\documentclass[conference]{new-aiaa}
\usepackage[utf8]{inputenc}
\usepackage{algpseudocode}
\usepackage{caption}
\usepackage{subcaption}
\usepackage{algorithm}
\usepackage{graphicx}
\usepackage{amsmath}
\usepackage[version=4]{mhchem}
\usepackage{siunitx}
\usepackage{longtable,tabularx}
\usepackage{xcolor}
\usepackage{booktabs,array}
\usepackage{multirow}
\usepackage{hyperref}
\usepackage{nomencl}
\usepackage{relsize}
\usepackage{gensymb}
\usepackage{graphicx}
\usepackage{array}
\usepackage{float}
\usepackage{hyperref}

\definecolor{blue}{rgb}{0.1 0.3 0.7}
\definecolor{green}{rgb}{0 0.4 0.2}
\definecolor{red}{rgb}{0.7 0 0}
\definecolor{yellow}{rgb}{0.9 0.7 0.1}

\hypersetup{
    colorlinks=true,
    linkcolor=black,
    filecolor=black,      
    urlcolor=black,
    citecolor=black
    }
\title{Reinforcement Learning-Based Closed-Loop Airfoil Flow Control}
\author[]{Qiong Liu\footnote{Assistant Professor. Department of Mechanical and Aerospace Engineering. }}
\author{Luis Javier Trujillo Corona\footnote{Graduate Student. Departent of Mechanical and Aerospace Engineering.}}
\author{Fangjun Shu\footnote{D.L. and A.G. Chapman Endowed Associate Professor. Department of Mechanical and Aerospace Engineering.}}
\author{Andreas Gross\footnote{Forrest Mooney Endowed Professor. Department of Mechanical and Aerospace Engineering.}}
\affil{New Mexico State University, NM 88003, USA}

\begin{document}

\maketitle
\begin{abstract}
\sffamily
We systematically investigated a reinforcement learning (RL)-based closed-loop active flow control strategy to enhance the lift-to-drag ratio of a wing section with an NLF(1)-0115 airfoil at an angle of attack $\alpha=5^\circ$. The effects of key control parameters—including actuation location, observed state, reward function, and control update interval—are evaluated at a chord-based Reynolds number of $Re=20,000$. Results show that all parameters significantly influence control performance, with the update interval playing a particularly critical role. Properly chosen update intervals introduce a broader spectrum of actuation frequencies, enabling more effective interactions with a wider range of flow structures and contributing to improved control effectiveness. The optimally trained RL controller is further evaluated in a three-dimensional numerical setup at the same Reynolds number. Actuation is applied using both spanwise-uniform and spanwise-varying control profiles. The results demonstrate that the pretrained controller, combined with a physics-informed spanwise distribution, achieves substantial performance gains. These findings extend the feasibility and scalability of a pretrained RL-based control strategy to more complex airfoil flows.
\end{abstract}

\section{Introduction}
Enhancing the aerodynamic performance of airfoils has long been a central pursuit in flow control, dating back to the early days of aviation~\cite{anderson2011ebook}. 
Engineers try to leverage flow control to reduce the takeoff and landing speed, to improve cruise performance, especially for low-Reynolds number applications, to increase maneuverability, and to improve safety. These objectives continue to motivate extensive research in both passive and active flow control designs.

Recent advances in computational capability~\cite{noack2011reduced} and sensor-actuator hardware enhancement~\cite{cattafesta2011actuators} have elevated expectations for sustainable high-performance operation across both on-design and off-design conditions. These demands challenge the effectiveness of passive flow control~\cite{cattafesta2008active}, which cannot be deactivated when not needed, and open-loop flow control techniques~\cite{joslin2009fundamentals}, which rely on pre-defined control laws that are decoupled from the undergoing flow state. In contrast, closed-loop flow control provides an adaptive and potentially more robust solution by actively adjusting control inputs based on real-time flow state estimates. Despite its promise, designing effective closed-loop controllers for complex flows remains a significant challenge.

An essential closed-loop flow controller consists of sensors, actuators and a control law to determine the control output. The synergy among these three components determines the control effectiveness and efficiency. Within such frameworks, selecting an appropriate control law often involves a trade-off between simplicity, accuracy and robustness. Reduced-order models~\cite{taira2017modal}, and linear feedback controllers~\cite{sipp2010dynamics,sipp2016linear} offer computational efficiency but may fail to incorporate nonlinear interactions essential for accurate control. In some instances~\cite{aastrom2021feedback}, these simplifications result in suboptimal or unstable performance, especially in the presence of small but dynamically significant nonlinear effects.

Reinforcement learning (RL) as one kind of machine learning~\cite{mnih2015human, schulman2017proximal}, has shown great promise for designing nonlinear control policies. RL has been widely exploited in multiple research areas, such as robotics~\cite{kober2013reinforcement}, gaming~\cite{silver2016mastering}, autonomous vehicles, and finance. Leveraging the nonlinear nature of neural networks, RL can learn control laws through interaction with complex flow environments, making it particularly attractive for closed-loop flow control. For instance, ~\citet{fan2020reinforcement} demonstrated the feasibility of using RL to learn an effective closed-loop control policy for a circular cylinder in wind tunnel experiments.

Previously, reinforcement learning-based (RL-based) flow control has been employed for the drag reduction of circular cylinder flows at low Reynolds numbers. 
\citet{rabault2019artificial} used RL algorithm for the active control of a cylinder wake by adjusting wall blowing and suction at two locations on the cylinder. The control policy achieved approximately 8\% of drag reduction for a very small flow control mass flow rate. Follow-up work by \citet{li2022reinforcement} considered a cylinder flow which is confined on both sides. \citet{li2022reinforcement} investigated how to embed physical information into the reinforcement learning-based control to obtain a more efficient, effective, and robust control. They used a reward function that incorporates the growth and decay of the perturbations based on linear stability theory~\cite{theofilis2003advances, theofilis2011global} to fully suppress vortex shedding in the wake region. This indicates that the integration of flow physics can improve the success of reinforcement learning-based flow control strategies. These early successes highlight the potential of physics-informed RL for flow control, though the nonlinearities of RL were not fully taken advantage of for these relatively canonical flow problems.

Nonlinear network control laws can outperform conventional control designs even when tackling complex flow problems. \citet{sonoda2022reinforcement} designed RL-based control strategies for reducing skin friction drag in a fully developed turbulent channel flow. The same flow control technique, blowing and suction at the wall, was employed using a RL control based on a linear and nonlinear network. The nonlinear network control law accomplished a significant 14\% greater drag reduction than the linear control law. The study revealed the potential of using reinforcement learning to obtain control laws with embedded nonlinearity. It suggests that RL can deliver optimal control when properly tuned.

More and more studies~\cite{vignon2023recent,ghraieb2021single} demonstrate the benefits of using RL for closed-loop flow control problems. The achievable control performance is largely determined by the open parameters, particularly regarding the choice of key hyperparamters. The locations of sensors and actuators have a significant impact on the control effectiveness, as in traditional control theory. Beyond that, the other open parameters, such as the reward function and the time interval for the interaction of the control agent with the flow, also have a critical impact on learning efficiency and optimal performance. An improperly scaled reward function can yield a poor balance between exploration and exploitation~\cite{ibrahim2024comprehensive}, the balance between trying new forcing and exploiting known good ones, leading to suboptimal polices. In this study, we systematically examine the effect of reward function scaling in the context of airfoil flow control.

Very few investigations have considered and discussed the effect of the update interval between control law and environment which is a critical parameter in flow control applications. ~\citet{li2022reinforcement} and \citet{rabault2019artificial} used an integer multiple of the natural shedding period, typically six to eight vortex shedding cycles, from the circular cylinder as an update interval. However, the influence of the update interval on RL performance has not been thoroughly explored. In our previous study~\cite{trujillo2024active}, we followed a similar heuristic strategy and took six to eight times of the primary oscillation period as the update interval. In the present study, we systematically vary the update interval and investigate the impact on the control performance. We examine the update interval together with the actuator control mechanism to uncover the underlying connection between the chosen actuator and the update interval. This investigation is critical for selecting the appropriate temporal resolutions in RL-based flow control (update interval, computational timestep in simulations, sampling rate in experiments, etc.) for achieving optimal performance. It also offers guidance for future applications across different flow configurations.

Another practical challenge in RL for flow control is data efficiency.
Data requirements for RL training can be challenging, particularly when they are based on high-fidelity three-dimensional simulations. A natural question arises: can a control policy trained for one flow condition generalize to others? Our previous study~\cite{trujillo2024active} demonstrated that an RL agent trained for a Reynolds number of $Re=5,000$ can perform effectively at a higher Reynolds number of $Re=10,000$, as long as the underlying flow physics are similar. This suggests that scalability of a well-designed control technique depends more on the incorporation of the correct flow mechanisms than on the particular architecture or precise numerical values. ~\citet{font2025deep} further supported this idea by using deep RL to control turbulent separation bubbles. They compared deep RL control effectiveness to classical periodic forcing. The study demonstrates that a deep RL-trained agent, even when trained using simulations on a computationally cheaper coarse grid, can outperform traditional periodic control in reducing turbulent separation in simulations on a finer grid. The deep RL agent achieves this by discovering and sustaining large-scale counter-rotating vortices through a wide range of frequencies. In other words, as long as the dominant flow structures are recognized in the low-fidelity simulations, the control agent will likely also be successful in high fidelity simulations. These findings highlight the feasibility of training control agents in simplified settings while still achieving high performance in more complex environments.

To further explore generalizability and reduce training costs, we test whether a two-dimensional RL-trained controller can be applied in a three-dimensional setting with the aid of physical information. Specifically, we examine whether insights and policies from well-understood 2D flows can serve as a starting point for testing in more realistic 3D flows. This addresses the critical question of scalability across both Reynolds number and spatial dimensionality. These investigations naturally raise additional questions: (1) How should the observed state be selected to balance observability and efficiency? and (2) How should actuators be placed in 2D simulations to maximize transferability to 3D configurations? These topics are addressed in detail in this study.

The remainder of the paper is organized as follows: Section~\ref{sec:Problem} describes the problem setup and computational framework. Section~\ref{sec:LST} provides a brief overview of the linear stability framework. Section~\ref{sec:RLC} introduces the reinforcement learning-based control strategy. The numerical setup validation and the uncontrolled flow field are described in Sections~\ref{sec:Valid} and~\ref{sec:Baseline}, respectively. Results for 2D RL training and control performance are presented in Section~\ref{sec:2dtrain}. Section~\ref{sec:3dtest} presents the 3D test control results. Section~\ref{sec:summary} summarizes key findings and discusses future directions.

\section{Problem Description and Methodology}\label{sec:Problem}

\subsection{Separated flow over wing section with NLF(1)-0115 airfoil}

Flow over a two-dimensional and three-dimensional wing section with natural laminar flow airfoil NLF(1)-0115 at an angle of attack of $\alpha=5\degree$ is examined. The flow problem is characterized by the geometry of the airfoil and the incoming flow conditions. The chord length of the airfoil is set to $c=1$. The airfoil maximum thickness is $15\% \cdot c$ at $0.441c$ and the maximum camber is $1.8\%\cdot c$ at $0.3c$. The airfoil coordinates were obtained from Ref.~\cite{Airfoiltool}. We refined the airfoil geometry around the leading edge to obtain smooth numerical simulation results. The refined airfoil geometry data is available at: \href{https://github.com/QiongFluid/Airfoil_NLF/blob/main/Refined_airfoil_NLF_AoA5.csv}{GitHub.com}.  

\begin{figure}
    \centering   
    \includegraphics[width=\textwidth]{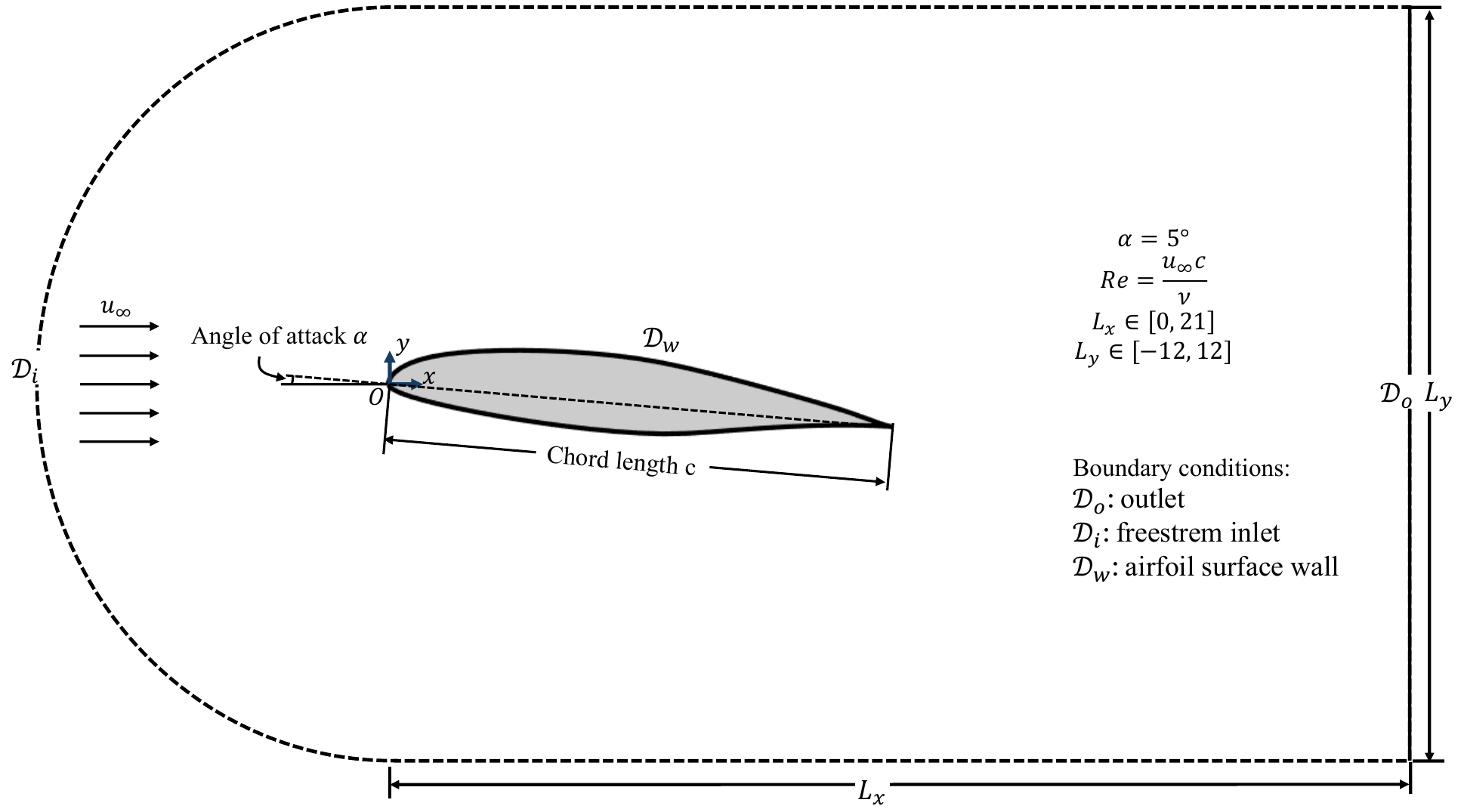}
    \caption{Description of flow problem (not-to-scale): Separated flow over NLF(1)-0115 airfoil at angle of attack $\alpha=5\degree$.
    \label{fig:airfoil_flow}}
\end{figure}

Figure~\ref{fig:airfoil_flow} shows the description for the flow over the airfoil at an angle of attack of $\alpha=5\degree$. 
A Cartesian coordinate system is defined with the streamwise $x$ and wall-normal $y$ axes, with the origin $(0,0)$ placed at the leading edge of the airfoil. The velocity components $\boldsymbol{u}(x, y, z, t)$ are in the direction of the Cartesian coordinates $(u, v, w)$ and inhomogeneous functions of the spatial coordinates and time. The Reynolds number, defined as $Re = u_\infty D/\nu$, is calculated based on the chord length $c$ and the freestream velocity $u_\infty$ where $\nu$ is the kinematic viscosity. Three Reynolds numbers, $Re=5,000$, 10,000, and 20,000, are examined by varying the kinematic viscosity.

\begin{figure}[H]
    \centering   
    \includegraphics[width=0.8\textwidth]{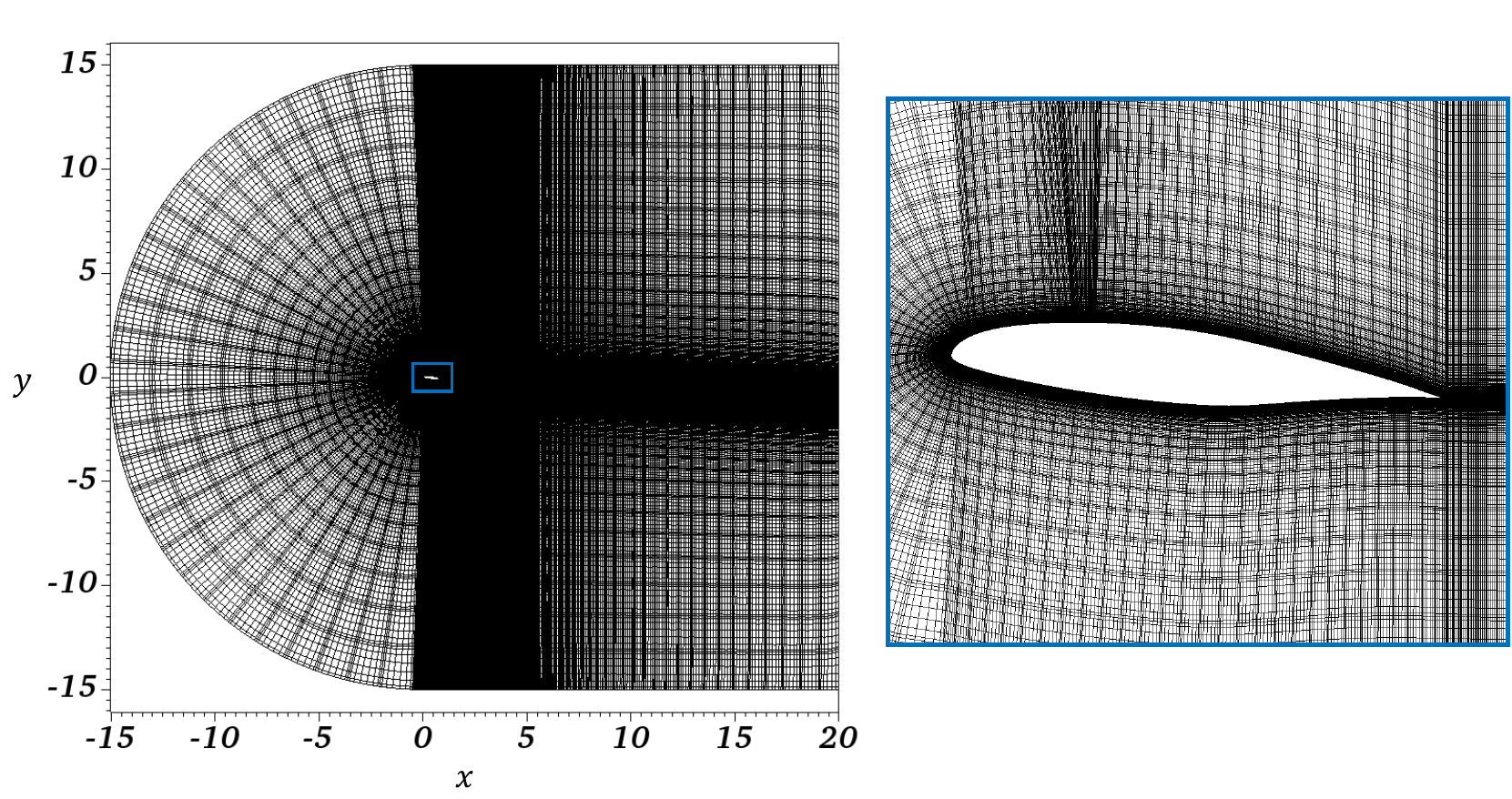}
    \caption{Full mesh and zoomed in view.}
    \label{fig:airfoil_mesh}
\end{figure}

2D direct numerical simulations (DNS) and 3D large-eddy simulations are performed using the incompressible higher-order-acurate spectral element solver Nek5000~\cite{Nek5000,liu2016linear}. A C-type computational domain is adapted with non-dimensional length $L_x\in[-15, 20]$ and height $L_y\in [-15, 15]$. For the 3D case, the spanwise extend is $-0.2 <L_z< 0.2$. Simulations of an airfoil section with laminar separation by ~\cite{Benton_Visbal_2019} and ~\cite{visbal2018analysis} using different spanwise domain extents revealed that a spanwise domain extent of 0.4 was sufficient to accurately capture all relevant flow physics. The 2D spectral element mesh comprises elements with $9698$ mirco elements. Using seven Gauss-Lobatto-Legendre collocation points quadrature~\cite{FISCHER2001265, trefethen2000spectral}, the resulting number of grid points is approximately $620,672$. The 3D mesh was obtained by extending the well-validated 2D mesh in the spanwise direction using a uniform distribution with 10 micro-elements. For the controlled cases, the mesh was refined around the actuation locations to accurately simulate the momentum flux. The full mesh and a zoomed in view are shown in figure~\ref{fig:airfoil_mesh}. The mesh was refined near the wall and flow control actuator.

The boundaries for the computational domain are here defined as $\mathcal{D}_i$, $\mathcal{D}_w$ and $\mathcal{D}_o$, representing the freestrem inlet, airfoil surface wall and outlet boundaries, respectively, as shown in Figure~\ref{fig:airfoil_flow}. We prescribed homogeneous Dirichlet conditions on the wall boundary $D_w$. A uniform freestream velocity profile $(u, v) = (u_\infty, 0)$ was adopted at the inflow boundary $D_i$; and Neumann conditions $(p{\bf{I}} - (1/Re) \nabla u) \cdot \vec{n} = 0$ were employed at the outflow boundary $D_o$, where $\vec{n}$ is the normal vector and {\bf I} is the unitary matrix. The velocity distribution across the forcing slot is detailed in Section~\ref{sec:actuation}.
The second-order-accurate backward finite difference method is employed for time integration. The dimensionless convective time is defined as $tu_\infty/c$. A fixed time step size $\Delta t=2e-4$ is adopted to ensure $CFL<1$ for all the cases.

\subsection{Unsteady actuation setup}\label{sec:actuation}
\begin{figure}
    \centering   
    \includegraphics[width=\textwidth]{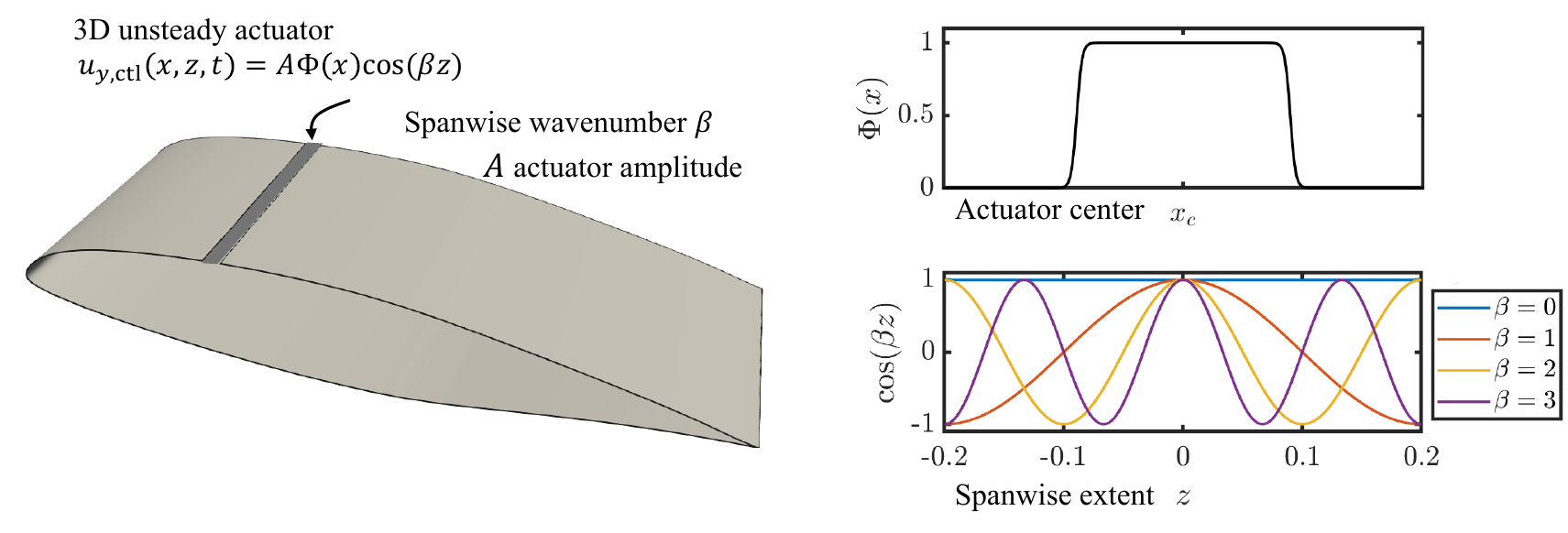}
    \caption{Schematic of unsteady actuator and velocity profile with spanwise variations. The value of the actuator amplitude $A$ is dictated by the reinforcement learning-based closed loop flow control. $\Phi(x)$ and $\cos(\beta z)$ are prescribed across the forcing slot.}
    \label{fig:3dactuator}
\end{figure}

We perform closed-loop airfoil flow control by introducing wall-normal blowing and suction across an actuation slot through a mass flux boundary condition. This is illustrated in figure~\ref{fig:3dactuator}. The wall-normal velocity distribution is
\begin{align}
    u_{y,\text{ctl}}(x,z,t)=A(t)\Phi(x)\cos(\beta_c z)
\end{align}
where the function $\Phi(x)=0.25\{1+\text{tanh}[a*(x-x_c+0.01)]\}\{1-\text{tanh}[a(x-x_c-0.01)]\}$ is employed to ensure a smooth transition from the no-slip and no-penetration wall conditions to the actuator conditions. Here, $a=2000$ and $x_c$ is the location of the center of the slot. The function $\cos(\beta_c z)$ describes the velocity profile in the spanwise direction with wavenumber $\beta_c$. For the 2D cases, $\beta_c=0$. For the 3D controlled cases, the wavenumber $\beta_c=\frac{2 n \pi }{L_z}$ is based on the width of the computational domain, where $n = 1, 2, 3\dots$. The two functions $\Phi(x)$ and $\cos(\beta_z z)$ are determined once the wavenumber is fixed. The amplitude $A(t)$ varied in time and was obtained from a reinforcement learning-based active flow control, which is introduced in section~\ref{sec:RLC}.

The control effectiveness is assessed through the aerodynamic efficiency $C_l/C_d$, where $C_l$ and $C_d$ are the lift and drag coefficients, respectively. The relative changes in drag, lift and aerodynamic efficiency are computed using
\begin{align*}
     \Delta \bar{C_d}=\frac{\bar{C}_\text{d,ctl}-\bar{C}_\text{d,unctl}}{\bar{C}_\text{d,unctl}},\quad
    \Delta \bar{C_l}=\frac{\bar{C}_\text{l,ctl}-\bar{C}_\text{l,unctl}}{\bar{C}_\text{l,unctl}}, \quad \text{and} \quad
    \Delta \overline{C_l/C_d}=\frac{(\overline{C_l/C_d})_\text{ctl}-(\overline{C_l/C_d})_\text{unctl}}{(\overline{C_l/C_d})_\text{unctl}},   
\end{align*}
where subscripts unctl and ctl indicate the uncontrolled and controlled cases, respectively.

\section{Linear Stability and Adjoint Analyses\label{sec:LST}}

Two-dimensional linear stability and adjoint analyses are applied to identify the intrinsic unsteady flow structures that inform sensor placement (observed state) and actuator location and design, both of which are considered indispensable for effective flow control. These analyses provide a physics-based foundation for guiding where to introduce actuation and observe the flow response.

We perform linear stability analysis following the classical theory~\cite{theofilis2011global,liu2016instability} and compute the corresponding adjoint modes~\cite{giannetti2007structural} using the spectral-element solver Nek5000~\cite{Nek5000, peplinski2013stability}. For completeness, we briefly summarize the theoretical formulation. Readers seeking a more comprehensive discussion are referred to Refs.~\cite{schmid2002stability, theofilis2003advances, schmid2014analysis}.

The linearized Navier-Stokes equations are obtained by decomposing the flow in a steady base flow $\boldsymbol{u(x)}$ and time-dependent small perturbations $\epsilon{\boldsymbol{u'(x)}}$ where $\epsilon<<1$. By retaining only the first-order terms (i.e., $O(\epsilon)$), the governing equations are obtained,  
\begin{align}\label{eqn:linNS}
    \frac{\partial \boldsymbol{u'}}{\partial t}=\mathcal{A}\boldsymbol{u'}
\end{align}
where the operator $\mathcal{A}$ is the Jacobian matrix of the right hand side of the linearized Navier–Stokes equations. 
Perturbation wave solutions of the form
\begin{align}\label{eqn:anzats}
    \boldsymbol{u'}(x,y,t)=\hat{\boldsymbol{u}}(x,y)\text{e}^{\lambda t} + \text{c.c.}
\end{align}
are assumed where $\lambda=\lambda_r+\text{i}\lambda_i$, $\lambda_i$ represents the oscillation frequency, $\lambda_r$ denotes the amplification/damping rate and c.c.\ is the complex conjugate. Substitution of Eq.\ (\ref{eqn:anzats}) into Eq.\ (\ref{eqn:linNS}) yields the eigenvalue problem
\begin{align}\label{eqn:eig}
    \lambda \hat{\boldsymbol{u}}=\mathcal{A} \hat{\boldsymbol{u}}.
\end{align}

The eigenvalues and eigenmodes characterize the global linear instability of the base flow. The spatial distribution of the direct modes informs the selection of sensor (observation) locations, where coherent structures are most pronounced.

The adjoint problem is formulated by solving the eigenvalue problem for the adjoint operator $\mathcal{A}^\ast$,
\begin{align}
   \lambda^\ast \hat{\boldsymbol{u}}^\ast=\mathcal{A}^\ast \hat{\boldsymbol{u}}^\ast, 
\end{align}
where $\hat{\boldsymbol{u}}^\ast$ denotes the adjoint eigenmodes. These adjoint structures identify the origin of the perturbations and thus guide the actuator placement.

For the numerical solution of the direct and adjoint problem, the same 2D domain and mesh as for the DNS are employed. The perturbation boundary conditions for the stability analysis match those of the DNS, except for the inflow condition $\mathcal{D}_i$, where $\hat{\boldsymbol{u}}=(0,0)$.

\section{Reinforcement Learning-Based Active Flow Control}\label{sec:RLC}

We designed a closed-loop flow control strategy based on an RL framework. The RL closed-loop flow control is based on five critical components---environment, RL agent, actuation (i.e., forcing), observed state and reward. The environment is defined by the numerical simulations of the flow over the airfoil section with unsteady forcing. The RL agent manipulates the environment via the flow control, i.e., the amplitude of the actuation. Then the change in the environment due to the forcing is fed back to the RL agent via the observed state. Simultaneously, the estimated reward based on the changed environment is provided to the RL agent as an assessment of the control effectiveness. Based on the estimated reward, the network weights are updated via back propagation. The five critical components establish a closed loop and synchronously improve the control performance. Ideally, the optimal control performance is reached within a few training episodes.

\begin{figure}[H]
    \centering   \includegraphics[width=\textwidth]{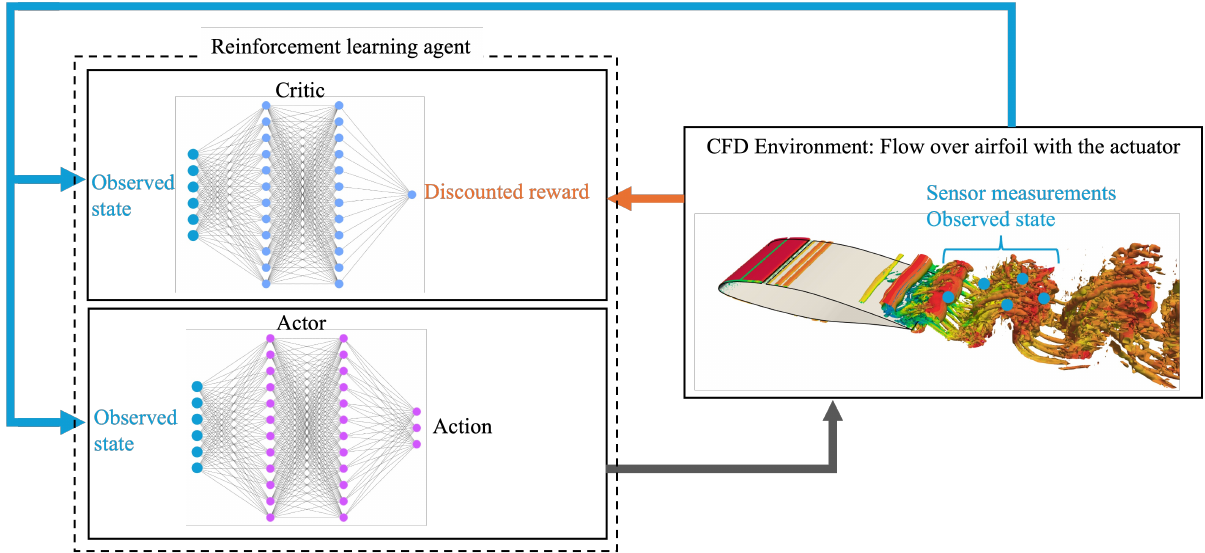}
    \caption{Reinforcement learning-based feedback loop for active airfoil flow control.}
    \label{fig:RL}
\end{figure}

Figure~\ref{fig:RL} shows a schematic of the RL-based flow control algorithm. We use the ``actor-critic" RL algorithm~\cite{Konda2003}. It contains two sets of multi-layer artificial neural networks. One of them, the Actor, learns a policy (a function that maps state to actuation) to determine which forcing amplitude is optimal in a given situation. The other, the Critic, evaluates the state after the flow control has been applied. The input layer for the Actor and Critic are the same, the observed state, while the output layers are different, the forcing amplitude and the predicted reward, respectively. The purpose of these two neural networks is to decide on a desired forcing amplitude that can maximize the accumulated reward. In this study, the architecture for both neural networks is the same and described by two dense layers of 512 fully connected neurons, plus input layers required to acquire data from the environment and an output layer for the forcing amplitude and the reward, respectively. The weights are updated using back-propagation for both neural networks. ReLU activation functions are used for the hidden layers with a fixed learning rate of 0.001.

The RL agent perceives the change in the environment through observed states. The locations of the observed state are critical for an effective control, as the actor learns a function that maps the state to the actuation. The problem of effectively observing the state is related to theory in control design that looks for effective sensor locations. For suppressing instabilities in an oscillator flow~\cite{sipp2016linear}, the sensor is typically placed where the most unstable global mode shows a maximum, as this ensures an effective sensing even in the presence of additive sensor noise. For the present results the sensors were placed such that the RL agent can recognize the global modes. We test three different sensors configurations. The sensors are either located in the wake or they are distributed over the airfoil surface. In contrast, the actuator is placed in a region where the flow is most receptive to the forcing, so that a minimal amount of actuation can have a significant effect on the flow. In the context of stability theory, this location corresponds to a large adjoint mode amplitude or, for input-output or resolvent analysis, a large forcing mode amplitude. This placement makes the control effective and efficient, as the forcing input can be amplified by underlying instabilities.

The definition of the reward function directly decides the control objective. RL agents try to maximize the total expected reward by repeatedly interacting with the environment. The RL agent adjusts the forcing amplitude by assessing the reward function. We decided on a reward function of $r=C_l/C_d-p$, with constant $p=10$ and 20 to check the impact on the learned optimal control policy. The reward function defines the exploration and exploitation relationship. The comparison and the effect will be discussed in Section~\ref{sec:reward}.

\begin{figure}
\centering
\includegraphics[width=0.9\textwidth]{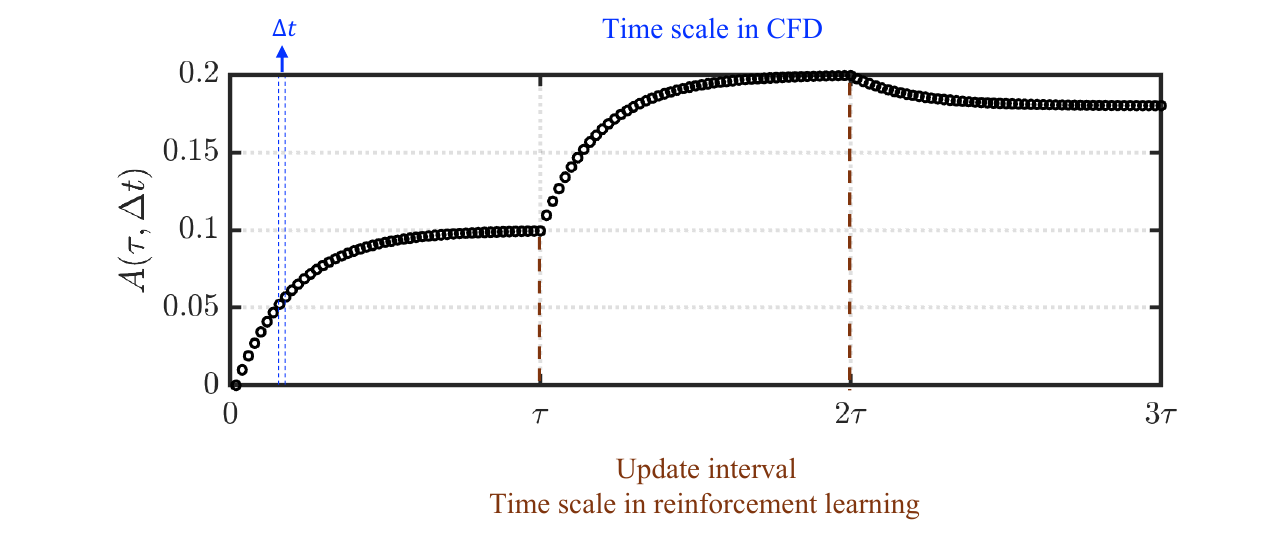}
\caption{Different time scales for CFD simulation and RL agent. Continuous update of action input over one action period $\tau$. $\Delta t$ is the simulation time step size.}
\label{fig:timescale}
\end{figure}

The agent was allowed to interact with the flow for a limited time after which the Critic observed the performance based on the reward function therefore facilitating control policy learning. More specifically, the agent was allowed to interact with the environment and update its forcing every hundred simulation time steps. Every $N$ steps of the updates are called one episode. Figure~\ref{fig:timescale} illustrates relevant time scales in the control design framework. The different time scales of the intrinsic fluid dynamics decide the interaction time scale. If the interaction time period is too short or too long, the learning will be inefficient. The interaction time remains an open parameter that requires examination to assess its effect on the control policy learning. A previous study for a cylinder flow~\cite{li2022reinforcement} suggests updating the RL forcing every $6-10\%$ of the vortex shedding period. A time scale that is either too short or too long may lead to failure of the training process.

Instead of abruptly updating the forcing amplitude with each newly learned value, the amplitude of the actuation updates iteratively in time. The jet actuation amplitude is updated using a first-order relaxation scheme:
\begin{align}
A(n+1) = A(n) + \beta \left( A_k - A(n) \right), \quad \text{for } \frac{(k-1)\tau}{\Delta t} \leq n < \frac{k\tau}{\Delta t}, \quad k = 1, \dots, N\\
A(0)=0 
\end{align}
where $A_k=A(k\tau)$ is the amplitude value selected by the RL agent at the beginning of the $k$-th update interval of duration $\tau$. The index 
$n$ denotes the CFD time step and $\Delta t$ is the CFD time step size. The under-relaxation factor is $\beta = 0.1$ ensures a gradual transition of the actuation amplitude over each control interval, mitigating abrupt changes that could lead to numerical instability. Figure~\ref{fig:timescale} illustrates this interative update over multiple RL forcing intervals, showing how the actuation transitions toward new targets set by the RL policy. To alleviate the adjustment effect on the control, the reward estimate is only evaluated over the last 10 time step, when the actuator amplitude has already converged to $A_\text{RL}(2\tau)$. Thus the selection of the under-relaxation factor will have a minimal effect of the control.

\section{Validation of Computational Setup}\label{sec:Valid}
\begin{figure}[H]
     \centering
     \begin{subfigure}[b]{0.48\textwidth}
         \centering
         \includegraphics[width=\textwidth]
         {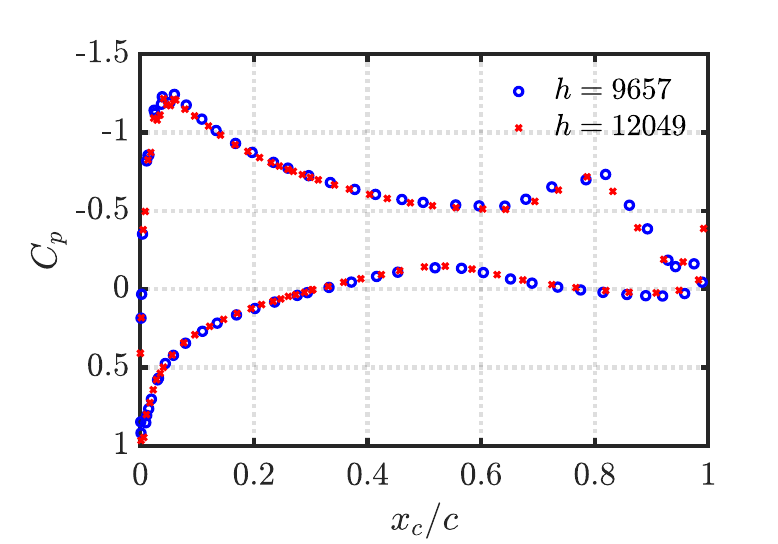}
         \caption{}
     \end{subfigure}
     \hfill
     \begin{subfigure}[b]{0.44\textwidth}
     \centering
    \includegraphics[width=\textwidth]{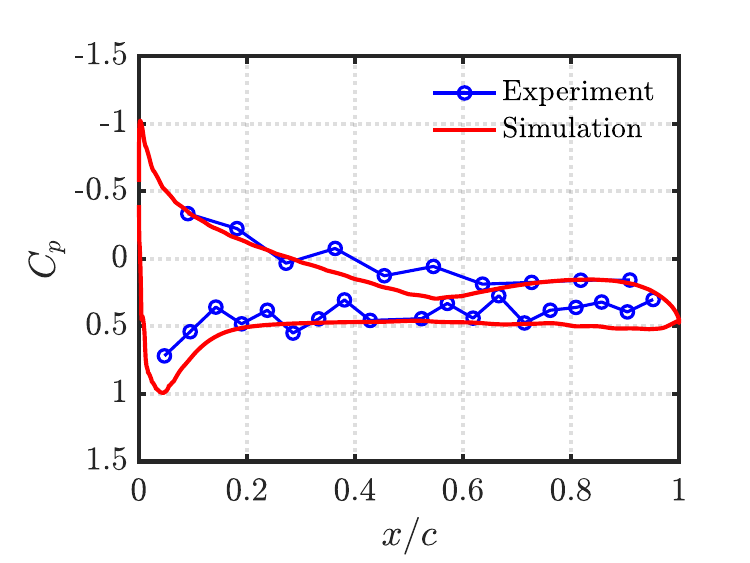}
      \caption{}
     \end{subfigure}
        \caption{Wall pressure coefficient over airfoil (a) 2D simulations for $\alpha=5^\circ$ and (b) 3D simulation and measurements for for $\alpha=10^\circ$.}
        \label{fig:mesh_vali}
\end{figure}

Mesh independence of the computational results was validated by increasing the number of micro-elements from $h = 9657$ to 12049, using the same polynomial order of 8. Figure~\ref{fig:mesh_vali}(a) shows the time-averaged wall-pressure distributions at $Re = 20{,}000$ and $\alpha = 5^\circ$. The good agreement of the time-averaged wall-pressure distributions indicates that the results are grid-independent. Based on this validation, we performed the simulations using the mesh with $h = 9657$. For $Re = 20{,}000$ and $\alpha = 10^\circ$, the wall pressure coefficient was compared with available experimental data (unpublished measurements in New Mexico State University wind tunnel), as shown in Figure~\ref{fig:mesh_vali}(b). The reasonable agreement validates the computational setup for the airfoil flow. Given the small dynamic pressures at the low Reynolds number, the pressure changes are minimal, which presents a challenge for accurate \( c_p \) measurements. The waviness observed in the experimental data is a result of measurement uncertainty. 

\section{Airfoil Flow Characteristics}\label{sec:Baseline}
Uncontrolled 2D and 3D simulations, referred to as baseline cases, were run for 200 convective time units ($t$), where time is normalized by the freestream velocity and chord length. These baseline simulations were continued until the flow reached a statistically stationary state and were then used for both initialization and comparison with the reinforcement learning-based control. The flow characteristics of the baseline cases are compared in the following section.

\subsection{Uncontrolled flow at $Re=5,000$, $10,000$ and $20,000$}

\begin{figure}[H]
     \centering
         \includegraphics[width=\textwidth]
         {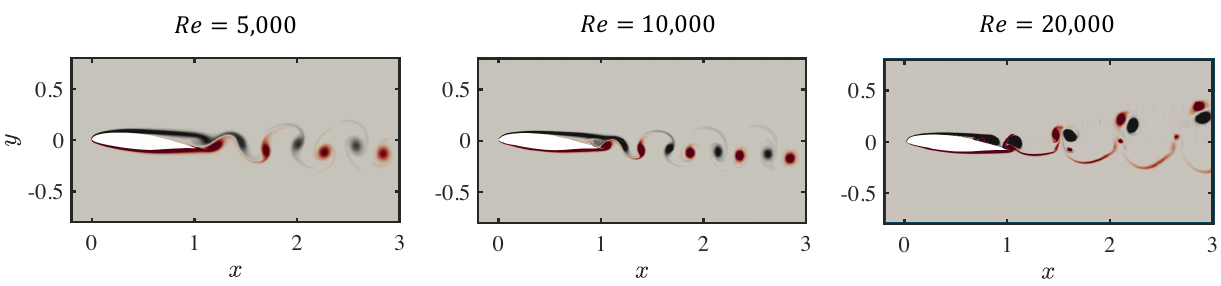}
         \caption{Instantaneous flow fields at $Re=5,000$, 10,000 and 20,000, visualized using $\omega_z$. The contour levels are between $\omega_z=-10$ (black) and $\omega_z=10$ (red).}
        \label{fig:basefl}
\end{figure}

Two-dimensional simulations were performed first.
The separated flow over the airfoil at $Re=5,000$, 10,000, and 20,000 was examined to elucidate variations in the flow that may influence the control efficacy. We analyzed and compared the instantaneous vorticity, the mean flow fields and skin-friction coefficient distributions for these three Reynolds numbers. The analysis aims to identify changes in the flow characteristics as the separation point moves upstream with increasing Reynolds number. More importantly, it serves to identify the potential Reynolds number validation range for a trained control agent.

Figure~\ref{fig:basefl} shows the instantaneous spanwise vorticity around the airfoil at three Reynolds numbers, $Re=5,000$, 10,000 and 20,000. These visualizations provide insight into the nature of the separated shear layer and the wake dynamics as the flow transitions with increasing Reynolds number. At $Re=5,000$, the flow exhibits a laminar trailing edge separation with low-amplitude, periodic vortex shedding in the wake.
The unsteady flow field resembles a von K\'arm\'an vortex street which is typical for bluff body wake flows.

For $Re=10,000$, the flow topology remains the same as for $Re=5,000$ but the vortex shedding sets in further upstream and the wake structures are spaced more closely.
At $Re=20,000$, the separated laminar boundary layer rolls up over the airfoil, suggesting shear-layer instability. Vortex shedding upstream of the trailing edge suggests an in the mean closed separation, as will be shown in the following section.
Compared to $Re=5,000$ and $10,000$, the wake vortex shedding pattern has changed.
 In this study, we develop RL-based flow control strategies for $Re=20,000$.  The lower Reynolds number cases are used to assess the scalability and robustness of the control across different flow regimes.

\begin{figure}
\centering
  \begin{subfigure}{\textwidth}
    \includegraphics[width=0.48\textwidth]{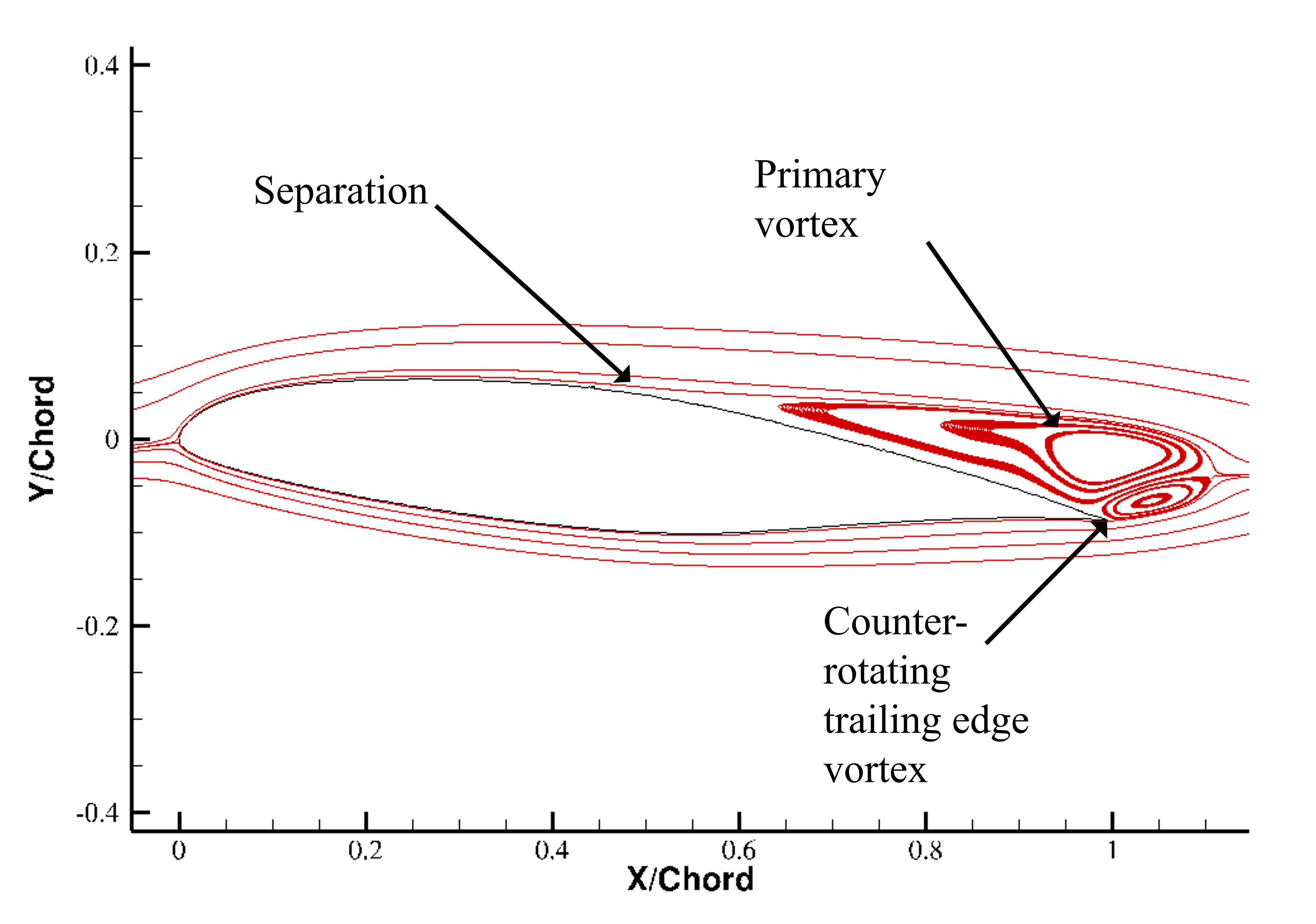}
    \includegraphics[width=0.48\textwidth]{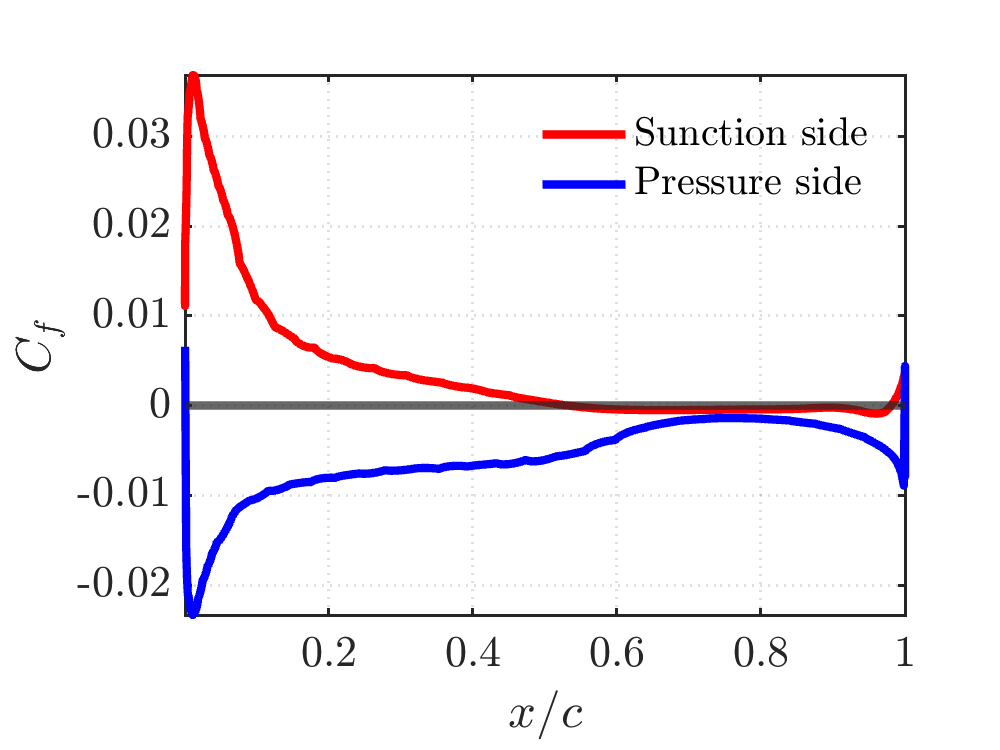}
    \caption{$Re=5,000$}
  \end{subfigure}
\hfill
  \hfill
    \begin{subfigure}{\textwidth}
    \includegraphics[width=0.48\textwidth]{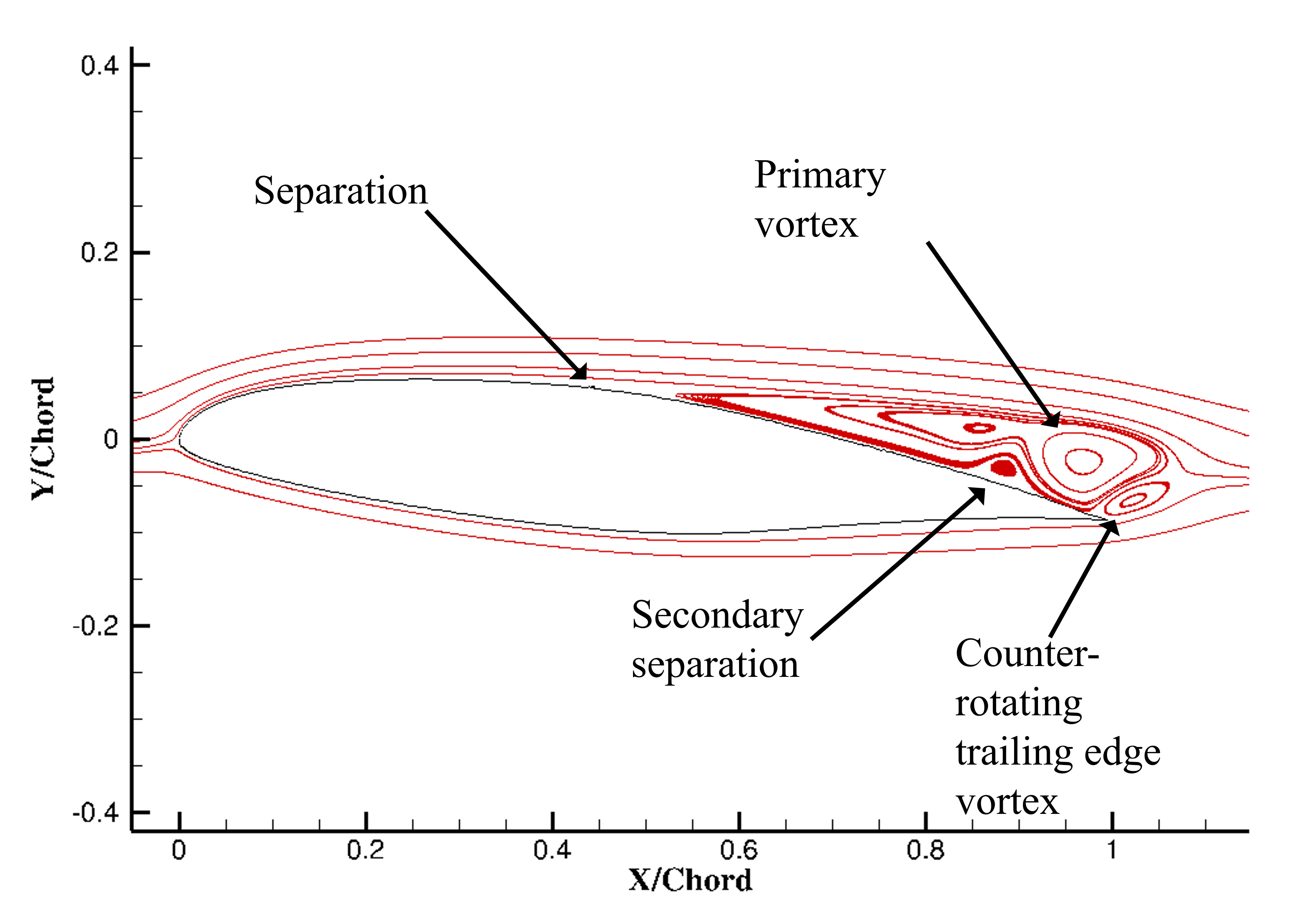}
    \includegraphics[width=0.48\textwidth]{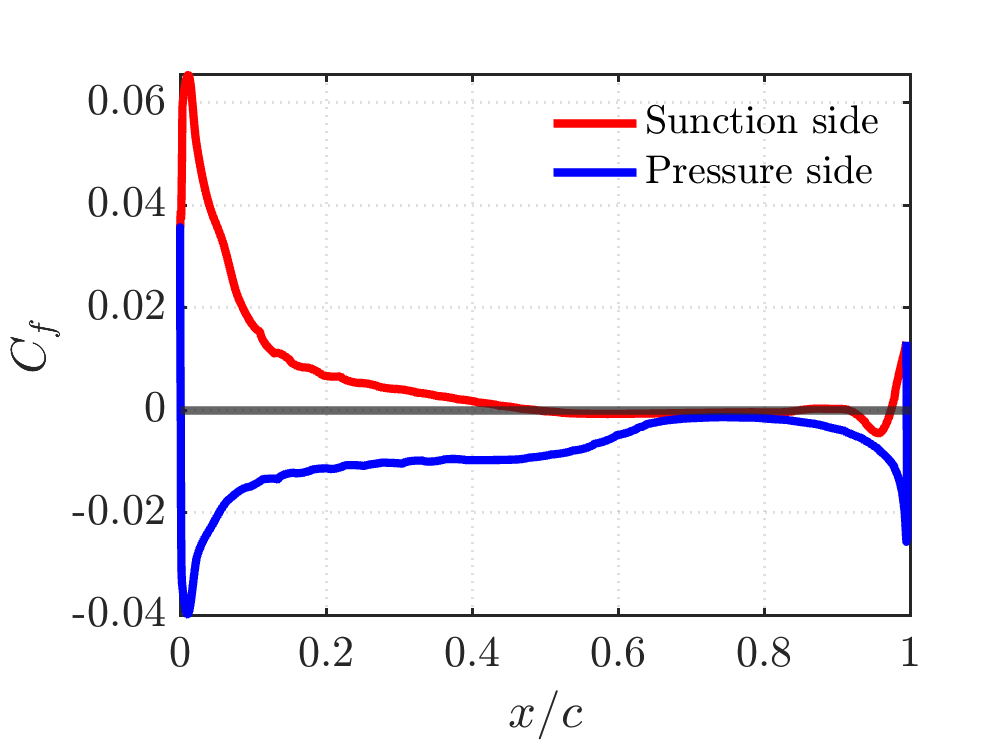}
    \caption{$Re=10,000$}
  \end{subfigure}
  \hfill
    \begin{subfigure}{\textwidth}
    \includegraphics[width=0.48\textwidth]{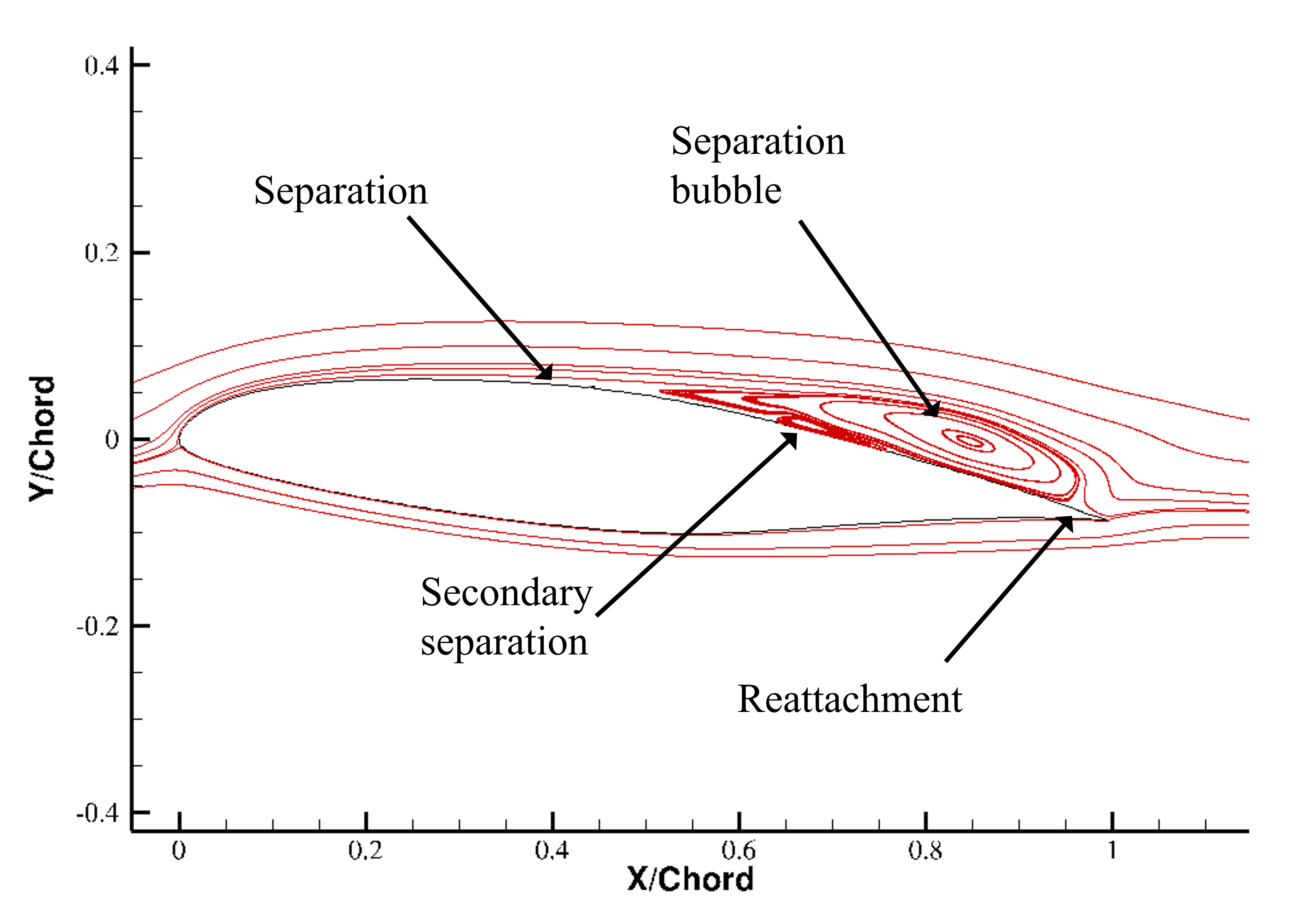}
    \includegraphics[width=0.48\textwidth]{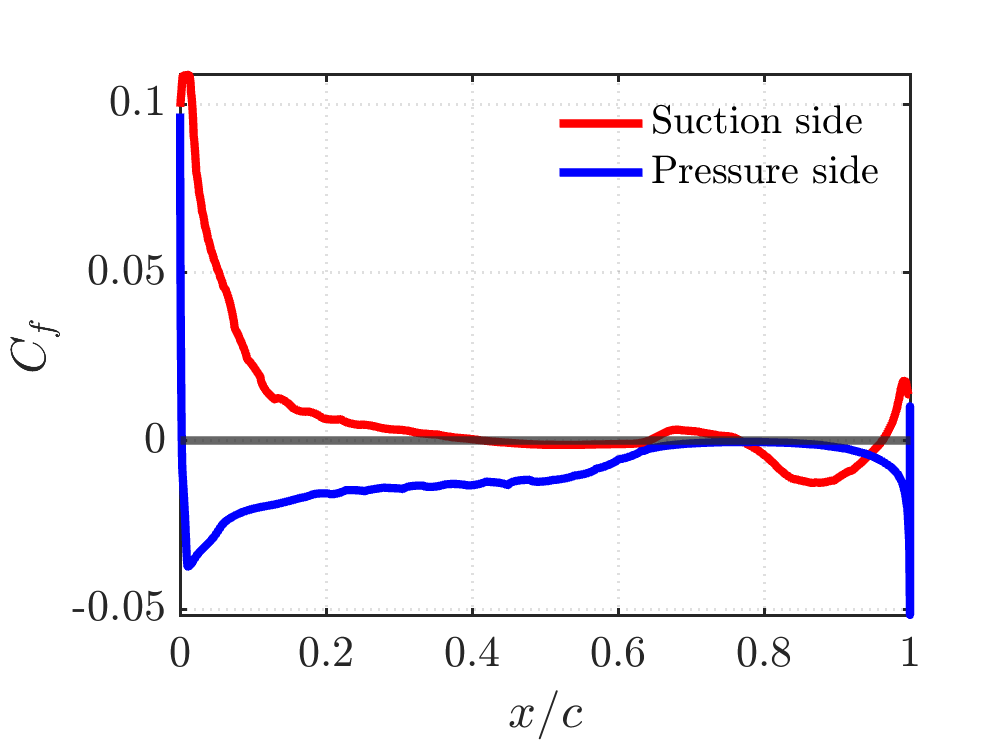}  
    \caption{$Re=20,000$}
  \end{subfigure}
  \caption{Mean flow streamlines and skin friction coefficient distributions for uncontrolled flow at $\alpha = 5\degree$.}
  \label{fig:fd}
\end{figure}

Figure~\ref{fig:fd}(a) shows mean-flow streamlines and skin-friction coefficient distributions for $Re=5,000$. 
The suction side boundary layer cannot negotiate the adverse pressure gradient and separates near mid-chord at $x/c=0.52$. The pressure side boundary layer separates at the trailing edge. In the mean flow, two counter-rotating vortices are observed. The larger primary vortex in the figure is rotating in the clockwise direction and the smaller trailing edge vortex is rotating in the counter-clockwise direction. This mean flow topology is common to wake flows. The skin-friction coefficient distribution helps pinpoint the mean separation location.
A sign-reversal of the skin-friction coefficient near the trailing edge is caused by the smaller counter-rotating trailing edge vortex.

As the Reynolds number is increased to 10,000, in the mean a notable secondary separation forms beneath the primary separation, as seen in Figure~\ref{fig:fd}(b). Compared to $Re=5,000$, the counter-rotating vortex at the trailing edge is smaller. The time-averaged skin-friction coefficient distribution reveals primary flow separation at $x/c=0.47$, which is slightly upstream compared to $Re=5,000$. Additionally, the secondary separation, which did not appear at the lower Reynolds number, is apparent for $0.85<x/c<0.91$. 
This secondary separation indicates that the shear layer roll-up occurs farther upstream and results in stronger rollers compared to $Re=5,000$.
Overall, the unsteady flow field remains governed by the wake instability.

Results for the time-averaged flow at $Re=20,000$ are shown in Figure~\ref{fig:fd}(c). Different from the lower Reynolds number cases, for this case separation has moved further upstream to $x/c=0.4$ and the flow reattaches in the mean at $x/c=0.92$ to form a laminar separation bubble.
The laminar separation bubble encloses an elongated secondary separation that spans from $x/c=0.62$ to $x/c=0.75$, as shown in Figure~\ref{fig:fd}(c). 
The secondary separation is also visible in the skin-friction coefficient distribution in Fig.~\ref{fig:fd}(c).
Also different from the $Re=5,000$ and $Re=10,000$ results, the small counter-rotating vortex is absent.
Different from $Re=5,000$ and $Re=10,000$, for $Re=20,000$ the unsteady flow is governed by a laminar separation bubble which is shedding co-rotating vortices.
This change in the unsteady flow physics has important implications for the control effectiveness as will be shown in the following section.

\subsection{Global direct and adjoint modes}

\begin{figure}
\centering
    \includegraphics[width=0.5\textwidth]{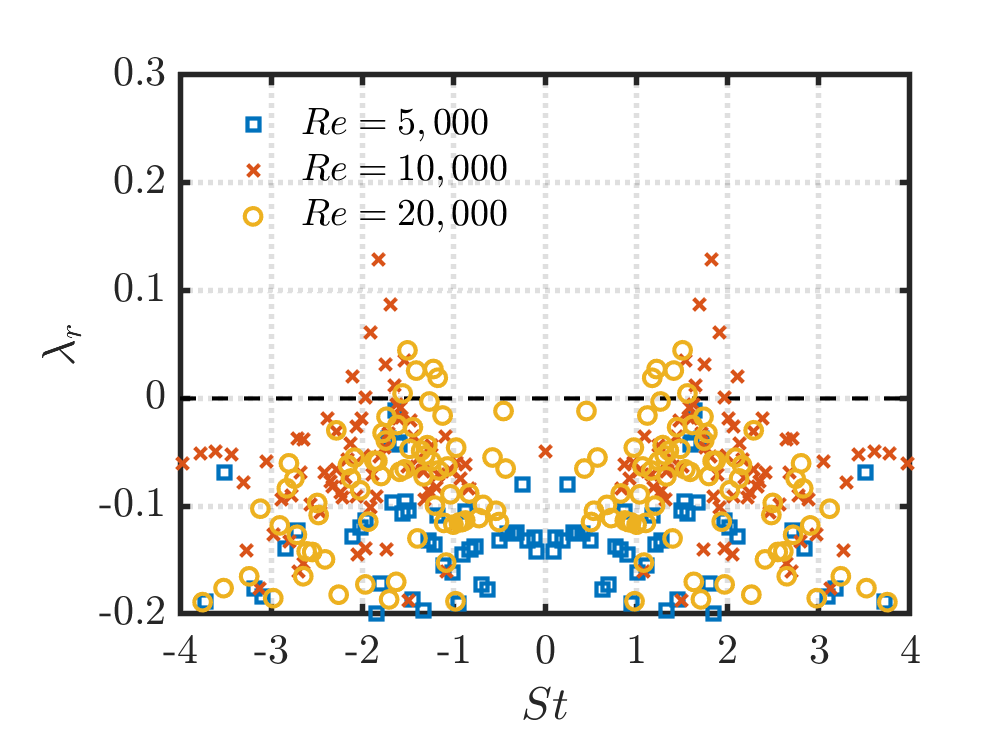}
    \caption{Spectra of the global stability modes at $Re=$ 5,000, 10,000 and 20,000 for an angle of attack $\alpha=5^\circ$. The black dashed line marked the neural stability threshold, $\lambda_r=0$.}  
  \label{fig:mode_spectra}
\end{figure}

Figure~\ref{fig:mode_spectra} shows spectra of global stability modes at $Re=5,000$, 10,000, and 20,000 for an angle of attack $\alpha=5^\circ$. The time-averaged flow fields served as base flows for the linear stability analysis. As the Reynolds number increases from $Re=5,000$ to 10,000, the least damped mode near $St\approx 1.65$ transitions to an unstable mode with a positive growth rate ($\lambda_r>0$), indicating that the time-averaged flow becomes unstable at this Reynolds number. At $Re=20,000$, the flow remains unstable. However, the growth rates are lower than for $Re=10,000$, suggesting that the establishment of the laminar separation bubble changes the dominant instability mechanism.

\begin{figure}
\centering
    \includegraphics[width=\textwidth]{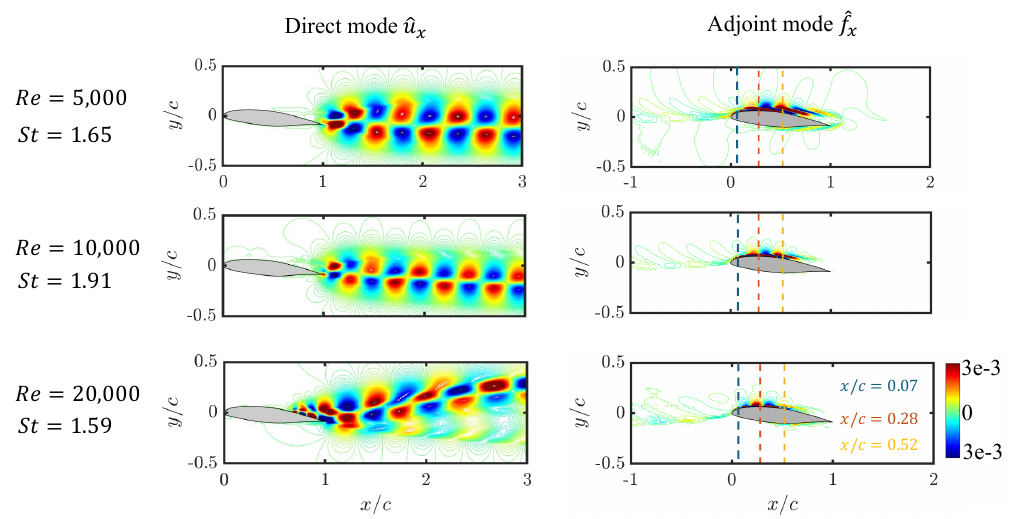}
    \caption{Most unstable global direct and adjoint modes for $Re = 5,000$, $10,000$, and $20,000$ and $\alpha = 5^\circ$. The left column shows the streamwise velocity component of the direct modes $\hat{u}_x$. The right column shows the corresponding adjoint modes $\hat{u}^\ast_x$. The colored dashed lines in the adjoint plots indicate the actuator locations: $x/c = 0.07$ (blue), $0.28$ (red) and $0.52$ (yellow).}  
  \label{fig:direct_adjoint}
\end{figure}

Figure~\ref{fig:direct_adjoint} shows the global direct and adjoint modes for the primary wake shedding frequencies at $Re=5,000$, 10,000, and 20,000 with $\alpha=5^\circ$. The time-averaged flow fields served as base flows for the analyses. Concerning the direct modes, the modes for $Re=5,000$ and 10,000 are qualitatively similar, and both resemble the classic wake mode~\cite{yeh2019resolvent}. The wake mode originates near the trailing edge and extends into the wake. The associated dominant frequency increases from $St=1.65$ to $1.91$ with Reynolds number increases from 5,000 to 10,000. 
At $Re=20,000$, however, the global mode structure is significantly different. The dominant mode, corresponding to a frequency of $St=1.59$, originates upstream near mid-chord and grows along the separated boundary layer. The shear layer mode transitions to a wake mode in the downstream direction. Different from the lower Reynolds number cases, the wake mode is tilted upward. Three sets of the observed states, including inside and outside of the direct mode will be examined and compared in the subsequent section.

The corresponding adjoint modes are shown on the right-hand side in figure~\ref{fig:direct_adjoint}. The adjoint modes reflect the flow's sensitivity to external perturbations.
They consistently exhibit large-amplitudes on the suction side of the airfoil prior to laminar separation.
With increasing Reynolds number, the adjoint mode amplitudes in the shear layer are reduced. 
The adjoint mode shapes indicate regions where the perturbations originate and are thus used to guide the actuator placement. Based on these analyses, three actuator locations are selected for the subsequent control studies. The location are $x/c=0.07$, 0.28 and 0.51, and are marked by dashed lines in the adjoint plots. The effect of the actuator location (upstream to mid-chord region) on the control effectiveness is evaluated in the next section.

\section{Reinforcement Learning-Based Flow Control at $Re=20,000$}\label{sec:2dtrain}
In this section, we investigate the effectiveness of RL-based flow control at a Reynolds number of $Re=20,000$, with a focus on the influence of key training parameters. Specifically, we analyze the effects of (i) the reward function $r$, (ii) the policy update interval $\tau$, (iii) the actuation location $x_c$, and (iv) the observed state $os$. These parameters are foundational to the design and success of RL-based controllers, and their selection often has strong connections to the underlying flow physics. By systematically varying these parameters, we aim to identify combinations that promote faster convergence of the RL-based flow control and an enhanced aerodynamic performance.

\subsection{Effect of reward function $r$}\label{sec:reward}

\begin{table}
\begin{minipage}{0.5\textwidth}
		\centering
		\includegraphics[width=\textwidth]{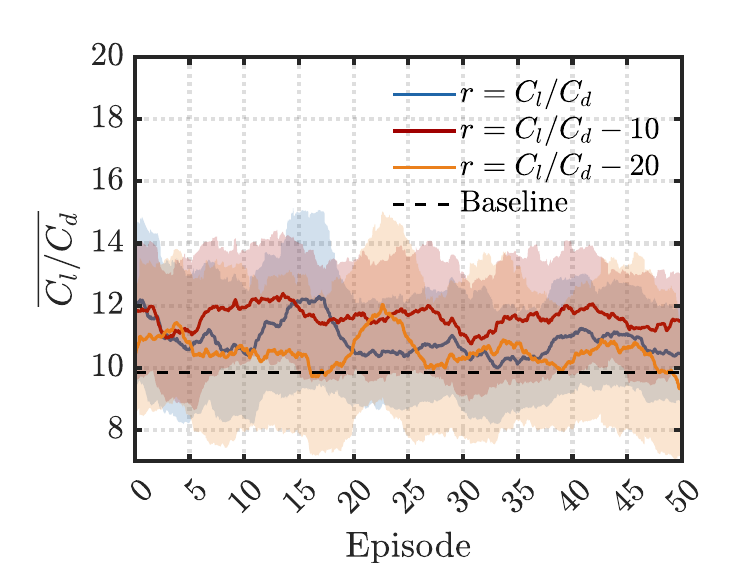}
		\captionof{figure}{Training histories for three reward functions at an identical update interval $t'=0.35$, observed state $os$ and actuation locations $x_c$.}
		\label{fig:history1}
	\end{minipage}\hfill
	\begin{minipage}{0.45\textwidth}
		\centering
		\begin{tabular}{lrrrr}
			\toprule
			$r$          & $\overline{C_l/C_d}$ & $\Delta(C_l/C_d)$ & $\overline{C_l}$& $\overline{C_d}$ \\
			\midrule
            $C_d/C_d$    & 10.539      & 4.5\%  &0.548 &0.052\\
			$C_l/C_d-10$ & 11.289      & 11.9\% &0.587 &0.052\\
			$C_l/C_d-20$ & 10.076      & -0.1\% &0.534 &0.053\\
            Baseline     & 10.085      & -      &0.716  & 0.071\\
			\bottomrule
		\end{tabular}
        \caption{Summary of aerodynamic efficiency for different reward functions.}
        \label{table:history1}
	\end{minipage}
\end{table}

The reward function is a central component in RL design, directly influencing the behavior and performance of the control agent. In the context of airfoil flow control, the objective is to improve aerodynamic efficiency, measured by the lift-to-drag ratio $C_l/C_d$. Accordingly, three different reward functions were considered: 
\begin{itemize}
    \item $r=C_l/C_d$: a direct representation of the performance metric.
    \item $r=C_l/C_d-10$: a shifted reward where values above 10 yield positive rewards.
    \item $r=C_l/C_d-20$: a more stringent criterion requiring $C_l/C_d>20$ for a positive reward.
\end{itemize}
These reward functions were tested using the same update interval $t'=0.35$. The aim was to explore how reward scaling and penalty influence the learning dynamics, particularly in terms of policy convergence, exploration and finally control performance.

Figure~\ref{fig:history1} shows running averages and standard deviations (over 200 time steps) of the aerodynamic efficiency $C_l/C_d$ over 50 training episodes. The x-axis represents the training episodes. The agent trained with a shift reward of $r=C_l/C_d-10$ consistently achieves superior aerodynamic efficiency, with a smoother and faster convergence compared to the other cases. In contrast, the $r=C_l/C_d-20$ configuration results in notably poorer performance, with the agent failing to reach the baseline efficiency (see Tab.~\ref{table:history1}). This degradation is attributed to the high penalty which shifts the reward signal predominantly into the negative range, discouraging exploration and hindering effective learning.

The performance differences are quantitatively summarized in Table~\ref{table:history1}(b). The best-performing case, $r=C_l/C_d-10$, achieves an $11.9\%$ improvement in aerodynamic efficiency relative to the baseline, whereas only a modest $4.5\%$ gain is obtained with the unbiased reward function. However, for $r=C_l/C_d-20$ the controller slightly under-performs compared to the baseline.

The observed behavior aligns with theoretical insights into actor-critic algorithms, where the reward signal acts analogously to a temperature parameter in an energy-based policy framework. When the reward scale is too small, the policy tends towards uniformity, limiting exploitation of learned behavior. Conversely, excessively large or consistently negative rewards can cause premature convergence to suboptimal deterministic polices, due to a lack of sufficient exploration. In practice, reward scaling is a critical hyperparamter. The present results demonstrate that a moderate negative shift in reward provides a good balance between exploration and exploitation. This tuning guides the agent effectively through the control space, resulting in both faster learning and improved asymptotic performance.

\subsection{Training with different update intervals $\tau$}

\begin{table}
\begin{minipage}{0.5\textwidth}
		\centering
		\includegraphics[width=\textwidth]{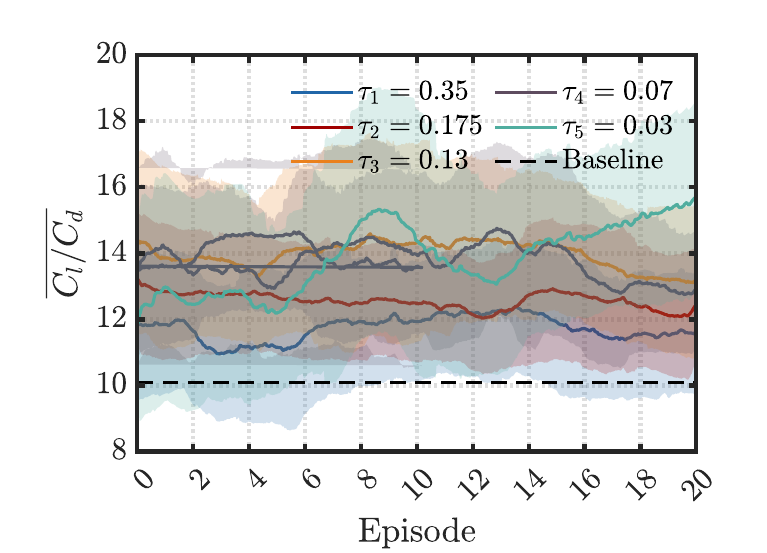}
		\captionof{figure}{Training histories of aerodynamic efficiency ($C_l/C_d$) for different update intervals $\tau$, using reward function $r=C_l/C_d-10$ and actuation location $x_c$=0.28. Solid lines represent moving averages computed over a 200 data points sliding window; shaded areas indicate one standard deviation.}
		\label{fig:history2}
	\end{minipage}\hfill
	\begin{minipage}{0.45\textwidth}
		\centering
		\begin{tabular}{lrrrr}
			\toprule
			$\tau$     &$\overline{C_l/C_d}$ & $\Delta(C_l/C_d)$ & $\overline{C_l}$& $\overline{C_d}$ \\
			\midrule
			$0.35$     & 11.289     & 11.9\% &0.587&0.052\\
            $0.175$    & 12.275     & 21.7\% &0.584&0.048\\
	     	$0.13$     & 13.370     & 32.6\% &0.615&0.046\\
			$0.07$     & 13.704     & 35.9\% &0.621&0.045\\
            $0.03$     & 13.063     & 29.5\% &0.627&0.048\\ 
            Baseline   & 10.085     & -      &0.716&0.071\\
			\bottomrule
		\end{tabular}
        		\caption{Summary of aerodynamic performance across different update intervals $\tau$.}
                		\label{table:history2}
	\end{minipage}
\end{table}

Having identified $r=C_l/C_d-10$ as a reward function that yields strong control performance, we next investigate the effect of the update interval $\tau$ on reinforcement learning-based flow control. The update interval determines how frequently the control policy adjusts the actuator amplitude during training and it is directly related to the actuation Strouhal number $St$. We tested five different update intervals, $\tau_1=0.35$, $\tau_2=0.175$, $\tau_3=0.13$, $\tau_4=0.07$ and $\tau_5=0.03$. The corresponding Strouhal numbers are $St_1=2.857$, $St_2=5.714$, $St_3=7.692$, $St_4=14.286$ and $St_5=33.333$, respectively. These values are provided for reference in evaluating the actuator response capabilities.

Figure~\ref{fig:history2} shows the aerodynamic efficiency training histories for each case. A moving average with a sliding window of 200 points was applied, and the shaded areas denote the standard deviation. All controlled cases outperform the baseline. As $\tau$ decreases from 0.35 to 0.07, the control performance improves significantly both in terms of convergence rate and final aerodynamic efficiency. The case with $\tau=0.07$ achieves the highest performance with rapid convergence and consistent performance throughout the training process. The case with $\tau=0.03$ demonstrates a different training trend. The aerodynamic efficiency lags during the first six episodes, then increases and fluctuates around a relatively high level. This suggests that extremely small update intervals may lead to unstable learning dynamics or make the control oversensitive to short-term flow fluctuations, thereby impacting learning stability. 
Overall, $\tau=0.07$ yields the highest efficiency and most consistent learning behavior, thus enabling robust learning and an effective control.

Table~\ref{table:history2} summarizes the performance metrics for each update interval compared with the baseline. The aerodynamic efficiency improves progressively with decreasing update interval $\tau$, reaching a peak for $\tau=0.07$. 
While the improvement for the case with $\tau=0.13$ is 32.6\%, a value that is similar to the value for the best-performing case, the results suggest that the effectiveness of the reinforcement learning control deteriorates away from $\tau=0.07$.

\begin{figure}
\centering
    \includegraphics[width=\textwidth]{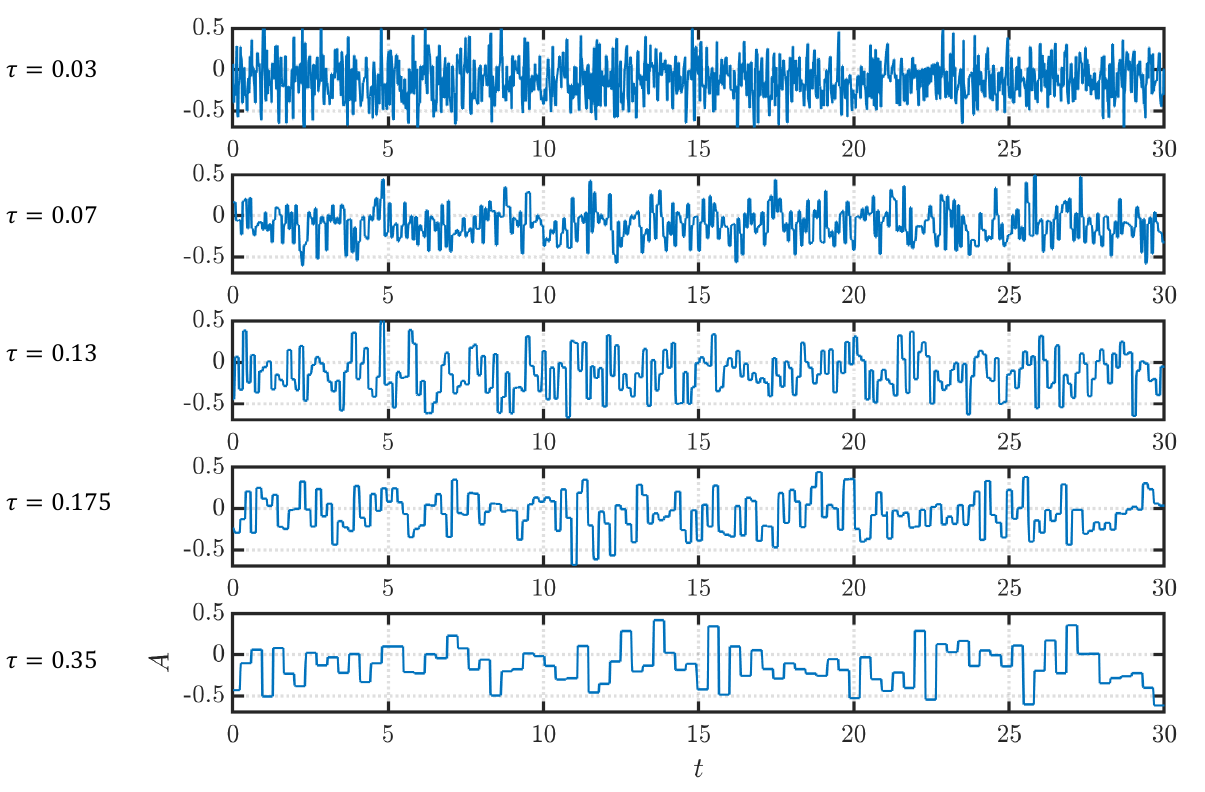}    
    \caption{Actuator amplitude time histories over a time horizon of 30 for different update intervals $\tau$.}   
  \label{fig:act_his}
\end{figure}

To gain further insights into the flow control behavior, figure~\ref{fig:act_his} shows the actuator amplitude histories over a time horizon of 30 units for each case. For large $\tau$, such as 0.35, the actuator signal appears as a clean square wave. As $\tau$ decreases, individual square waves become less visually discernible, and the actuation time histories transition into a more broadband signal. Although individual square waves become harder to distinguish visually when $\tau$ is lowered, zoomed-in views (not shown) confirm the presence of square waves for all cases. 
We hypothesize that for smaller update intervals, the controller introduces a broader range of actuation frequencies, potentially allowing for a more effective interaction with a wider range of flow structures, thus contributing to the control effectiveness.

\begin{figure}
\centering
    \includegraphics[width=\textwidth]{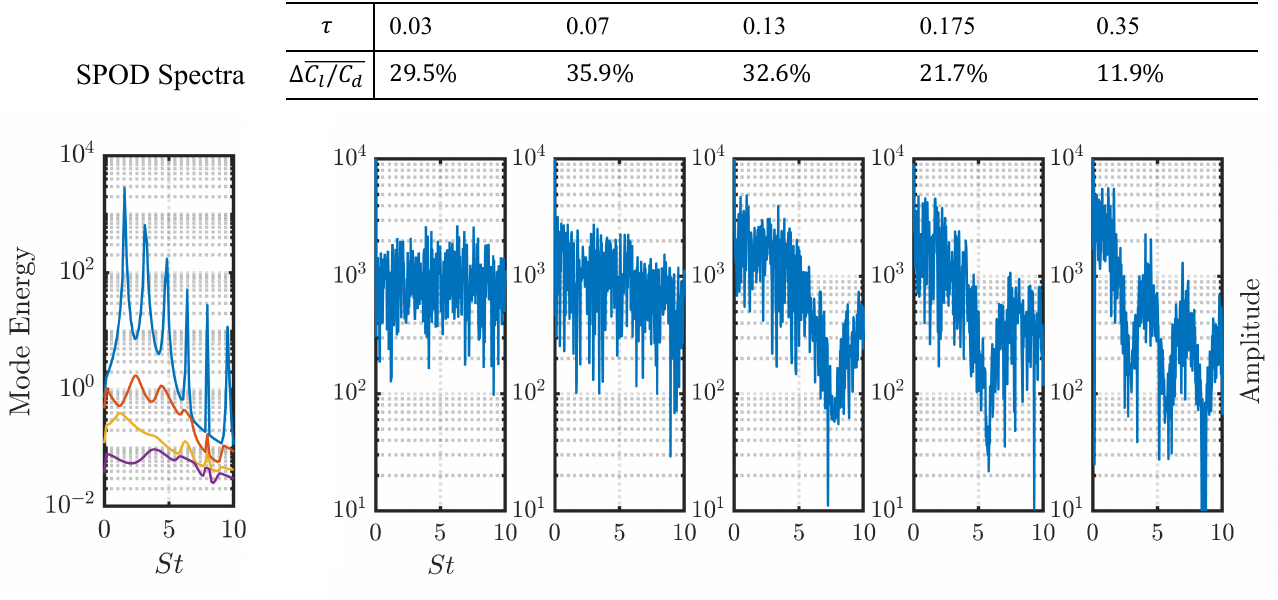}    
    \caption{Left: SPOD spectra for uncontrolled flow (ranked according to energy as blue, red, orange, and purple). Right: Forcing amplitude frequency spectra for controlled flow with different update intervals, $\tau$.}     
  \label{fig:act_st}
\end{figure}

To examine this hypothesis, fast Fourier transforms of the actuator amplitude time histories were computed and are compared in Figure~\ref{fig:act_st}. 
We also conducted Spectral Proper Orthogonal Decomposition~\cite{schmidt2020guide} to identify energy spectra for the uncontrolled flow.
In the uncontrolled flow, the frequencies associated with the dominant flow structures are concentrated at discrete tones. While for all controlled cases, the actuation spectra are broadband. For $\tau = 0.03$ (i.e., the shortest update interval), the actuator frequency content spans approximately $0 < St < 10$.
As the $\tau$ value increases (e.g., $\tau = 0.13$, 0.175, and 0.35), the spectral amplitude drops off for $St > 5$.
For the best-performing cases ($\tau = 0.07$ and 0.13), the spectral content within $0 < St < 5$ remains relatively even and high in amplitude.
These results provide a new perspective on selecting the update interval for reinforcement learning-based flow control across a wide range of flow configurations.

\subsection{Training with different actuation locations $x_c$}
To evaluate the influence of the actuator placement on the control effectiveness, three streamwise actuation locations were tested, $x_c=0.07$, 0.28 and 0.51. These locations were selected based on the adjoint mode distribution shown in figure~\ref{fig:direct_adjoint}, which highlights regions of high flow receptivity to actuation. 

Figure~\ref{fig:his_act} shows training histories for each actuator position, with aerodynamic efficiency tracked over 20 episodes. A moving average with a sliding window of 200 data points is applied to each case, and the shaded regions represent the corresponding standard deviation, indicating training variability. Among the three configurations, actuation at $x_c=0.28$ yields the hightest aerodynamic efficiency, achieving a $32.6\%$ improvement over the baseline. In contrast, shifting the actuator downstream to $x_c=0.51$ reduces the control effectiveness, resulting in a smaller improvement of $23.5\%$. This case also exhibits larger fluctuations in performance during training, suggesting a less stable policy convergence.

When the actuator is placed upstream at $x_c=0.07$, the control not only fails to improve performance but also significantly degrades it, with a 42.1\% reduction in $C_l/C_d$ relative to the baseline. Interestingly, the shaded area for this case is narrow, reflecting low variability in the training history. This suggests that the reinforcement learning agent quickly converges to a stable, but ineffective, control policy. The consistently poor reward signal likely discourages exploration, leading the agent to settle into a suboptimal strategy early in training. From a flow physics perspective, actuation near the leading edge may cause early boundary layer transition or laminar separation, both of which would deteriorate aerodynamic performance.

Table~\ref{table:sum_act} summarizes the aerodynamic performance metrics for each actuation location. The table reports mean values of the lift coefficient, drag coefficient, and aerodynamic efficiency obtained over 20 training episodes. Relative improvements are calculated with respect to the baseline. Actuation at $x_c=0.28$ produces the greatest improvement of $32.6\%$, while upstream actuation at $x_c=0.07$ significantly degrades performance, highlighting the importance of a physics-informed placement of the actuator.

\begin{table}
\begin{minipage}{0.5\textwidth}
		\centering
		\includegraphics[width=\textwidth]{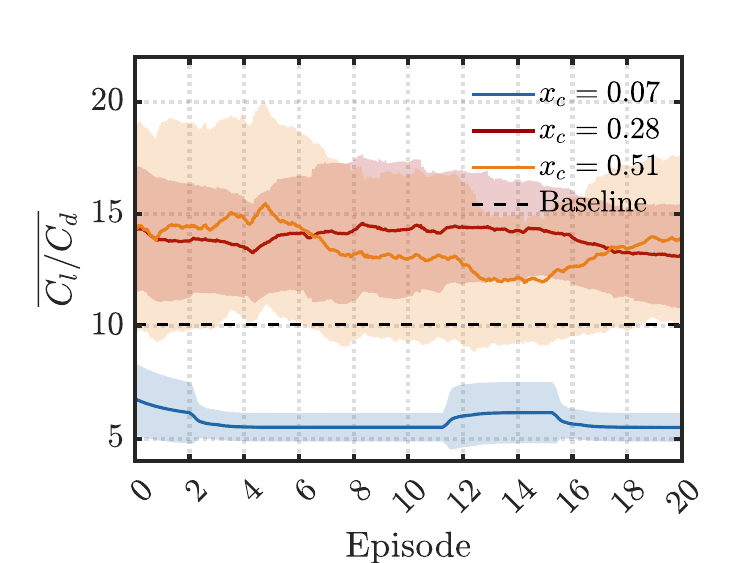}
		\captionof{figure}{Training histories of aerodynamic efficiency for three actuation locations $x_c=0.07$ (blue), $x_c=0.28$ (orange) and $x_c=0.51$ (red) using reward function $r=C_l/C_d-10$ and update interval $\tau=0.13$.}
		\label{fig:his_act}
	\end{minipage}\hfill
	\begin{minipage}{0.45\textwidth}
		\centering
		\begin{tabular}{lrrrr}
			\toprule
			$x_c$      & $\overline{C_l/C_d}$ & $\Delta(C_l/C_d)$ & $\overline{C_l}$& $\overline{C_d}$ \\
			\midrule
            $0.07$   & 5.844       & -42.1\% &0.374&0.064\\
			$0.28$    & 13.370     & 32.6\%  &0.615&0.046\\
            $0.51$   & 12.457      & 23.5\%  &0.573&0.046\\

            Baseline & 10.085        & - & 0.716& 0.071\\
			\bottomrule
		\end{tabular}
        		\caption{Summary of aerodynamic performance for different actuator locations.}
                		\label{table:sum_act}
	\end{minipage}
\end{table}

\subsection{Training with different observed states $os$}
\begin{figure}
     \centering
     \begin{subfigure}[b]{0.48\textwidth}
         \centering
         \includegraphics[width=\textwidth]{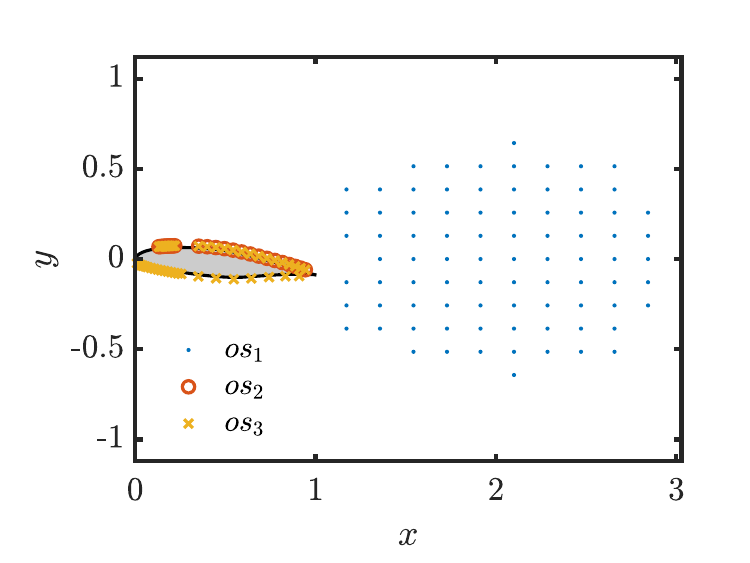}
         \caption{Observed state}
     \end{subfigure}
     \hfill
     \begin{subfigure}[b]{0.48\textwidth}
         \centering
         \includegraphics[width=\textwidth]{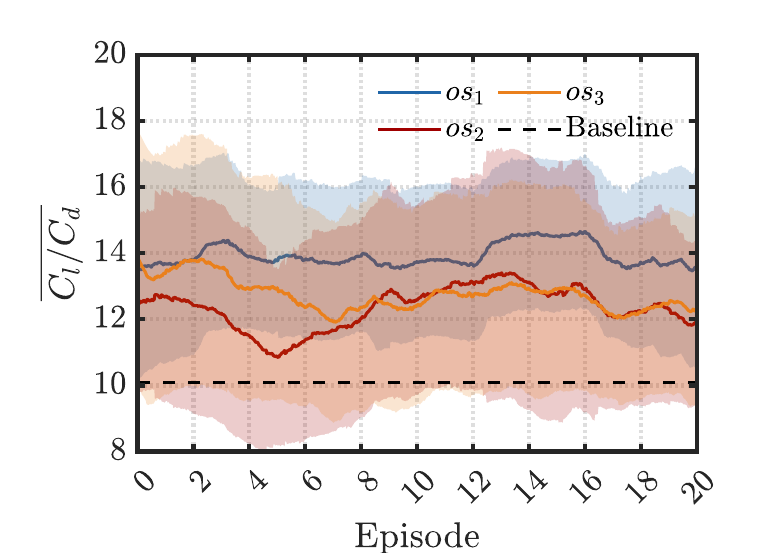}
         \caption{Lift}
     \end{subfigure}
        \caption{(a) Sensor layouts for three observed state configurations: $os_1$ (blue) includes 86 velocity sensors that are aligned according to the wake global mode structures; $os_2$ (orange) places 27 pressure sensors along the suction side; $os_3$ (red) expands $os_2$ to 51 sensors across both the suction and pressure sides. (b) Training history of aerodynamic efficiency for each observed state. Solid lines show the moving average over a 200-data points sliding window, shaded areas indicate one standard deviation, All cases use the same training parameters with update interval $\tau=0.13$ and reward function $r=C_l/C_d-10$.}
    \label{fig:os}
\end{figure}

In flow control, optimal sensor placement is often guided by the structure of the dominant adjoint global mode, which highlights regions of maximum receptivity to control inputs. However, in practical applications, full state observation is rarely feasible, and only partial state information is typically available. Therefore, it is essential to assess control effectiveness under limited observation conditions. Based on the global mode structures shown in figure~\ref{fig:direct_adjoint}, we define three different observed state configurations, as shown in figure~\ref{fig:os}(a).
$os_1$ includes 86 velocity sensors in the wake region. For this configuration the high amplitude structures of the global mode are covered by the sensors. $os_2$ consists of 27 pressure sensors along the suction side of the airfoil.
This configuration is more representative of practical applications where of-body measurements cannot easily be obtained.
$os_3$ expands this setup to 51 pressure sensors distributed across both the suction and pressure surfaces. These configurations are used to examine the impact of the state observation on the control performance.

Figure~\ref{fig:os}(b) shows the training histories for the different observed state configurations over 20 training episodes. The aerodynamic efficiency, quantified by the lift-to-drag ratio $C_l/C_d$, is evaluated over the training episodes using a moving average with a sliding window of 200 data points. The shaded areas indicate the corresponding moving standard deviation, reflecting the variability in the control performance. All three configurations were trained using identical settings: the update interval was set to $\tau=0.13$, the reward function was defined as $r=C_l/C_d-10$, and the initial flow state was identical. Each case was trained for 20 episodes. Among the three cases, $os_1$ (blue line) consistently yielded higher $C_l/C_d$ ratios, demonstrating superior control performance compared to $os_2$ (orange line) and $os_3$ (red line). This suggests that sensor placement in the wake region, where key flow structures develop, enables a more effective control. All controlled cases showed significant improvement over the baseline, indicated by the black dashed line.

Before the eighth episode, $os_2$ showed the lowest improvement. After episode 8, both $os_2$ and $os_3$ converged to similar levels of aerodynamic efficiency. Notably, $os_1$ showed greater variability in early episodes, indicating an initial exploration phase for the control agent. This is a characteristic behavior in reinforcement learning, where the agent explores a wide range of actions before gradually shifting toward exploitation of effective strategies.
Table \ref{table:history3} summarizes the overall aerodynamic performance improvements for each observed state configuration. $os_1$ achieved the highest improvement, with a $32.6\%$ increase in aerodynamic efficiency. $os2$ and $os_3$ yielded comparable gains of $14.7\%$ and $17.8\%$, respectively, for 20 training episodes. 

\begin{table}[H]
		\centering
		\begin{tabular}{lrrrr}
			\toprule
			$os$      & $\overline{C_l/C_d}$ & $\Delta(C_l/C_d)$ & $\overline{C_l}$& $\overline{C_d}$ \\
			\midrule
	$1, (u, v)$    & 13.370     & 32.6\%  &0.615&0.046\\
    $2, p$   & 11.571      & 14.7\%  &0.567&0.049\\
    $3, p$   & 11.875      & 17.8\%  &0.570&0.048\\
            Baseline & 10.085        & - & 0.716& 0.071\\
		\bottomrule
		\end{tabular}
        \caption{Summary of aerodynamic efficiency improvements for different observed state configurations.}
        \label{table:history3}
\end{table}

\section{Testing of Optimal Control Agent in 3D Simulations for $Re=20,000$}\label{sec:3dtest}
To evaluate the generalization of the trained controller, we test the optimal 2D RL control agent (trained using $r=C_/C_d-10$, $\tau=0.07$, $x_c=0.28$ and $os_1$) in a fully three-dimensional (3D) flow environment at $Re=20,000$. Except for the observed state, the 2D simulation setup is extended to a spanwise-periodic 3D configuration. The trained control policy is reused without retraining, and no trial-and-error data are required during this 3D test phase. To explore the effect of a spanwise variation of the actuation, we superimpose prescribed sinusoidal wavenumbers $\beta=0$, 1, 2 and 3 on the spanwise forcing amplitude. Here, $\beta=0$ corresponds to a spanwise-uniform control, while $\beta>0$ corresponds to a 3D variation of the forcing amplitude. The control effectiveness is compared to assess the 3D cases.

\begin{figure}[H]
     \centering
     \begin{subfigure}[b]{0.46\textwidth}
         \centering
         \includegraphics[width=\textwidth]{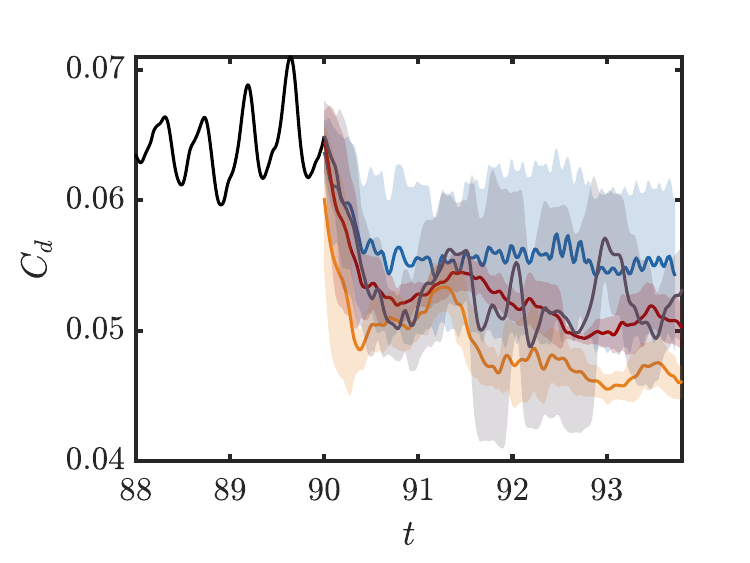}
         \caption{Drag}
     \end{subfigure}
     \hfill
     \begin{subfigure}[b]{0.46\textwidth}
         \centering
         \includegraphics[width=\textwidth]{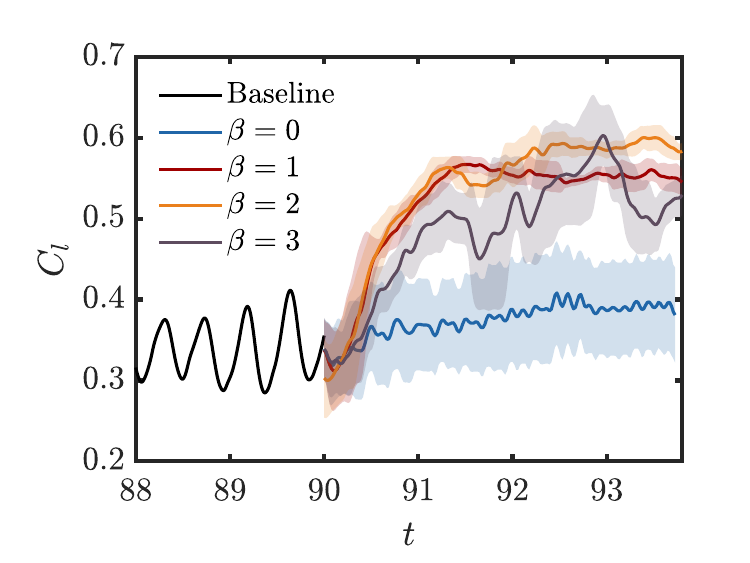}
         \caption{Lift}
     \end{subfigure}
        \caption{(a) Drag coefficient and (b) lift coefficient for controlled cases with different $\beta$ at $Re=20,000$. The data for $t>91$ are averaged to obtain the time-averaged aerodynamic efficiency.}
    \label{fig:test_Re20k}
\end{figure}

Figure~\ref{fig:test_Re20k} shows time histories of the drag and lift coefficient for the 3D controlled cases with different spanwise wavenumber. Time-averaged values and standard deviations were computed using a sliding window of 20 samples. 
The RL-based control strategy significantly improves aerodynamic performance across all cases. Lift increases while drag decreases compared to the baseline. Cases with $\beta=1$ and $\beta=2$ yield the most substantial improvements, with narrower fluctuation ranges and higher aerodynamic efficiency. In contrast, the $\beta=3$ case exhibits a less robust behavior with stronger fluctuations.

Table~\ref{tab:test3d} summarizes the time-averaged ($t>91$) lift and drag for the controlled cases. For all controlled cases a substantial enhancement in aerodynamic efficiency is achieved despite the fact that the control policy was trained in a 2D context. Among the different spanwise wavenumbers, the control policy combined with spanwise wavenumber $\beta = 1$ and 2 outperforms the others. These results confirm that a control policy trained in 2D can be effectively deployed in 3D environments, especially when combined with appropriately tuned spanwise actuation profiles.

\begin{table}[H]
    \centering
    \begin{tabular}{ccrrcc}
    \toprule
       Cases & & $\overline{C_l/C_d}$ & $\Delta(C_l/C_d)$ & $\overline{C_l}$& $\overline{C_d}$\\
        \midrule
       Baseline & & 5.277        & - & 0.343& 0.065\\ 
       &&&&&\\
       \multirow{4}{9em}{Controlled case} & $\beta=0$  &6.786 & 28.6\%&0.380 &0.056\\
       &$\beta=1$ & 10.434 & 97.7\% & 0.553 &0.052\\
       &$\beta=2$ & 12.000 & 127.4\%&  0.576& 0.048 \\
       &$\beta=3$ & 9.830  &  86.3\%&  0.521& 0.053 \\
       \bottomrule
\end{tabular}
\caption{Aerodynamic efficiency for 3D baseline and controlled cases. The statistically averaged quantities are obtained for $t>91$, as shown in figure~\ref{fig:test_Re20k}.}
\label{tab:test3d}
\end{table}

\begin{figure}
     \centering
     \includegraphics[width=0.7\textwidth, trim={0 0cm 0 0},clip]
         {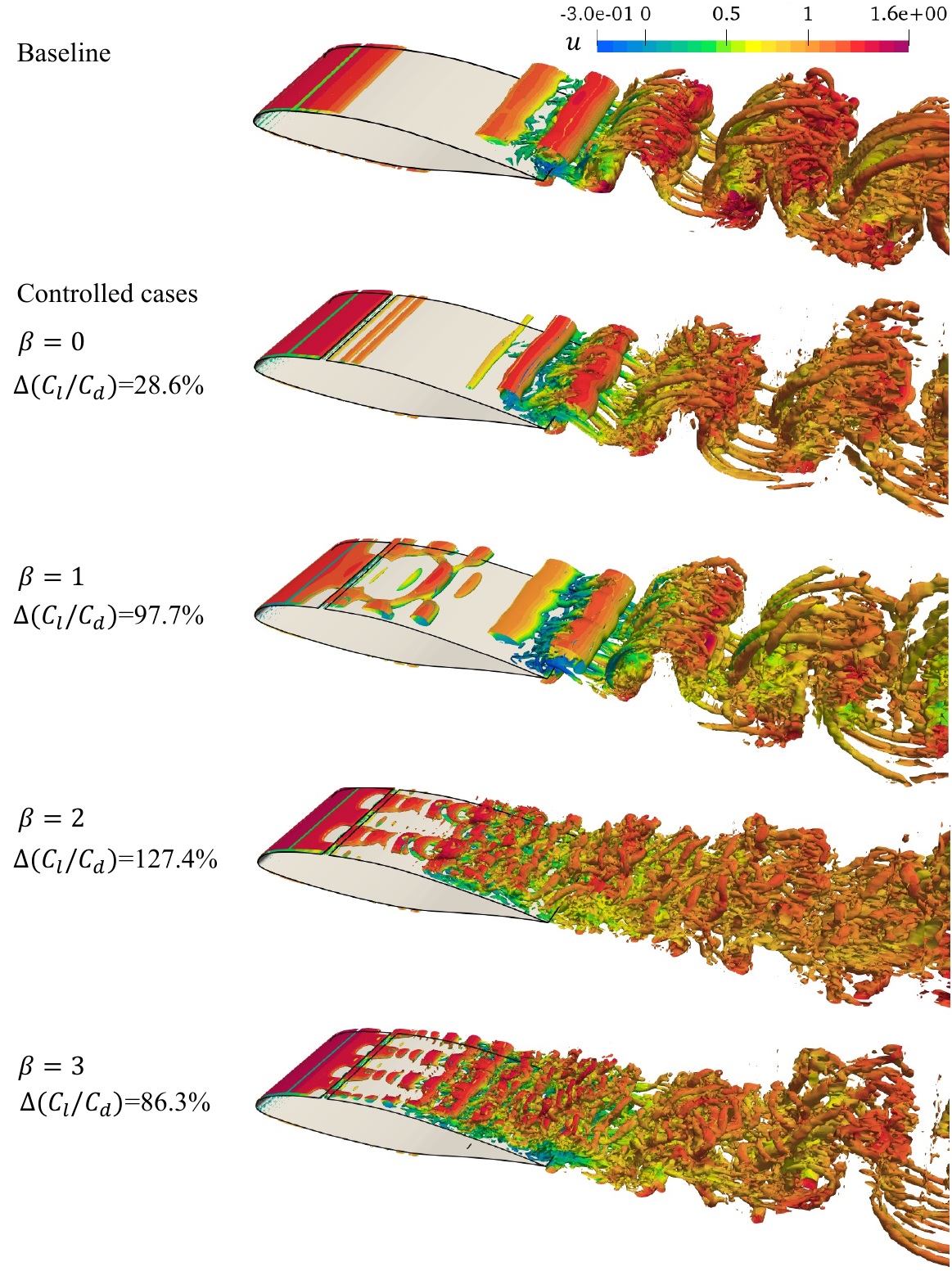}
        \caption{Instantaneous vortical structures for the baseline and controlled flows at $Re=20,000$, visualized using Q-criterion $Q=30$ and colored with streamwise velocity $u$.}
    \label{fig:q_criterion_test_20k}
\end{figure}

Figure~\ref{fig:q_criterion_test_20k} shows instantaneous vortical structures using the $Q$-criterion, colored by streamwise velocity $u$. The topmost figure represents the baseline case, while the subsequent plots correspond to controlled cases with spanwise wavenumbers of $\beta=0$, 1, 2, and 3, respectively. For the uncontrolled case, flow separation on the suction side is significant, characterized by strong shear layer roll-ups and wake vortex shedding. For example, two comparably strong roll-up structures with large spanwise coherence can be observed near the trailing edge. Downstream of the trailing edge, as the flow transitions to turbulence the structures slowly loose their spanwise coherence.

When applying two dimensional control with a spanwise wavenumber $\beta=0$, the strength and streamwise spacing of the shear-layer structures are reduced. Since the control is applied uniformly across the span, it primarily acts to stabilize the flow and weaken the vortex strength. Similar to the discussion in \citet{embacher2014direct} and \citet{benton2018high}, who showed that appropriate 2D forcing can delay laminar turbulent transition, breakdown appears delayed compared to the uncontrolled flow.
When the spanwise wavenumber is increased to $\beta=1$, vortical structures are discerned near the actuator with a noticeable spanwise variation. The spanwise perturbations are not amplified and the roll-up is similar to the uncontrolled and 2D controlled cases.
Nevertheless, the lift is increased and the drag is reduced compared to the uncontrolled case. 
Similar to the uncontrolled flow and the case with 2D forcing ($\beta=0$), the wake resembles a von K\'arm\'an vortex street.

When the spanwise wavenumber is further increased to $\beta=2$, the controlled flow is characterized by the break-up of the spanwise coherent structure on the suction side. The break-up of the vortices in the spanwise direction improves 3D mixing and suppresses flow separation near mid-chord. As a result of the earlier transition and reduced separation, von K\'arm\'an vortex shedding is suppressed and the wake appears fully turbulence. The drag is further reduced compared to the previous cases and reaches a minimum. Similar vortical structures are observed for the controlled case with spanwise wavenumber of $\beta=3$. Spanwise vortices are induced near the actuation slot, causing 3D mixing from mid-chord. 
Compared to $\beta=2$, the flow separates turbulent near the trailing edge. 
Further downstream, the wake transitions back to von K\'arm\'an vortex shedding. Even so, the actuation still achieves a significant aerodynamic efficiency gain compared to the uncontrolled flow.

The 3D results demonstrate that a control policy trained entirely in 2D can effectively improve aerodynamic efficiency when applied to 3D flows with superimposed spanwise wavenumbers. All tested cases show noticeable enhancements in aerodynamic performance, with the most significant improvement observed for $\beta = 1$ and $2$. The controlled case with $\beta = 0$ primarily acts to laminarize the flow and weaken the vortex strength. In contrast, the case with $\beta = 2$ promotes spanwise mixing and early laminar turbulent transition, yielding an effective separation suppression. These results highlight the robustness of the 2D-trained control agent and its feasibility for effectively controlling 3D flows with superimposed spanwise variation of the forcing amplitude.

\section{Conclusions}\label{sec:summary}
We developed a reinforcement learning-based closed-loop active flow control strategy to improve the aerodynamic efficiency of an NLF(1)-0115 airfoil at a Reynolds number of 20,000 and an angle of attack of $\alpha=5^\circ$. The control objective was to maximize the lift-to-drag ratio through wall-normal blowing and suction, with a control policy trained by interacting with a high-fidelity CFD environment.

We systematically investigated the effects of critical design parameters, including the actuation location, observed states, reward function and the update interval between the RL agent and the flow environment. Results showed that the update interval plays a significant role. An optimally chosen value ($\tau=0.07$) introduced broadband forcing and enhanced control performance. Similarly, physics-informed placement of actuators --guided by adjoint mode distributions-- greatly improved control efficiency. It suggests the importance of integrating flow physics into RL-based controller design.

We evaluated the generalization of the 2D-trained control agent in a 3D spanwise-periodic airfoil flow. Without additional training, the optimal 2D control policy, when combined with a prescribed spanwise variation, achieved significant control performance in 3D simulations. Specifically, controlled cases with spanwise wavenumbers of $\beta=1$ and $\beta=2$ yielded the highest aerodynamic performance, confirming the scalability and efficiency of the proposed RL control strategy. Furthermore, this study highlights the benefits of incorporating flow physics into the RL training. The use of linear stability and adjoint mode analysis provided essential insights into the most receptive flow regions for actuation and observation. This approach not only improved control performance but also reduced training cost and improved data efficiency.

In summary, this study demonstrates the feasibility and effectiveness of physics-informed RL control strategies for laminar separation control. The control agent trained in a 2D environment generalizes well to more complex 3D flows, offering a promising direction for future applications in real-world flow control. Future efforts will focus on higher Reynolds numbers, assessing robustness under uncertainties, as well as turbulent separation control.

\section*{Acknowledgments}
The authors acknowledge the support from the US Air Force Office of Scientific Research (Grant Number: FA9550-24-1-0069. Program Manager: Dr. Gregg Abate). Computational resources were provided through TACC Long-term tape Archival Storage (Ranch) and Purdue Anvil CPU under project numbers PHY230144 and PHY240233.

\bibliography{refRGD, sample, RL, ctl}

\end{document}